\documentclass[english,12pt]{article}
\pdfoutput=1
\usepackage{lmodern}

\usepackage[T1]{fontenc}
\usepackage[latin9]{inputenc}
\usepackage{array}
\usepackage{float}
\usepackage{booktabs}
\usepackage{mathrsfs}
\usepackage{multirow}
\usepackage{amsmath}
\usepackage{amssymb}
\usepackage{graphicx}

\makeatletter

\providecommand{\tabularnewline}{\\}

\newcommand{\lyxaddress}[1]{
	\par {\raggedright #1
	\vspace{1.4em}
	\noindent\par}
}

\usepackage{cancel}
\usepackage{mathrsfs}   %\mathscr{L}
\usepackage{slashed}     %\slashed{p}
\usepackage{bbold}  % \mathbb{1} for the identity matrix
\usepackage{url}
\usepackage{graphicx}
\usepackage[colorlinks=true,linkcolor=redLinks,citecolor=greenLinks,urlcolor=redLinks, pdfborder={0 0 1}]{hyperref}
\usepackage{xcolor}
\usepackage{framed}
\usepackage[numbers,sort&compress]{natbib}
\usepackage{amsmath}
\usepackage{ytableau}
\usepackage{microtype}

\usepackage{titlesec}

\allowdisplaybreaks

\colorlet{shadecolor}{gray!15}

\definecolor{greenLinks}{rgb}{0, 0.6, 0} 
\definecolor{blueLinks}{rgb}{0, 0, 0.6}
\definecolor{redLinks}{rgb}{0.6, 0, 0}
\definecolor{tempText}{rgb}{0.55, 0.10,0.67}
\definecolor{eprintLinks}{rgb}{0.4, 0.4, 0.4}
\definecolor{journalLinks}{rgb}{0.6, 0, 0}

\titleformat{\section}[block]{\color{black}\Large\bfseries\filcenter}{\thetitle.\;}{0em}{}
\titleformat{\subsection}[block]{\color{black}\large\bfseries\filcenter}{}{0em}{}

\newcommand{\MYhref}[3][redLinks]{\href{#2}{\color{#1}{#3}}}%

\usepackage{multirow}
\let\orig@Hy@EveryPageAnchor\Hy@EveryPageAnchor
\def\Hy@EveryPageAnchor{%
    \begingroup
    \hypersetup{pdfview=Fit}%
    \orig@Hy@EveryPageAnchor
    \endgroup
}

\let\oldFootnote\footnote
\newcommand\nextToken\relax

\renewcommand\footnote[1]{%
    \oldFootnote{#1}\futurelet\nextToken\isFootnote}

\newcommand\isFootnote{%
    \ifx\footnote\nextToken\textsuperscript{,}\fi}

\definecolor{myPurple}{RGB}{128,0,182}

\makeatother

\usepackage{babel}
\begin{document}
\title{Enumerating the operators of an effective field theory}
\author{Renato M. Fonseca\date{}}
\maketitle

\lyxaddress{\begin{center}
{\Large{}\vspace{-0.5cm}}Institute of Particle and Nuclear Physics\\
Faculty of Mathematics and Physics, Charles University,\\
V Holešovi\v{c}kách 2, 18000 Prague 8, Czech Republic\\
~\\
Email: fonseca@ipnp.mff.cuni.cz
\par\end{center}}
\begin{abstract}
Until recently little was known about the high-dimensional operators
of the standard model effective field theory (SMEFT). However, in
the past few years the number of these operators has been counted
up to mass dimension 15 using techniques involving the Hilbert series.
In this work I will show how to perform the same counting with a
different method. This alternative
approach makes it possible to cross-check results (it confirms the
SMEFT numbers), but it also provides some more information on the
operators beyond just counting their number. The considerations made
here apply equally well to any other model besides SMEFT and, with
this purpose in mind, they were implemented in a computer code.
\end{abstract}

\section{Introduction}

It is sometimes useful to consider interactions which are allowed
by symmetry, even if they are not renormalizable. Seen as effective
interactions, they can be used to study the effects of new, 
higher-energy physics in a model-independent way.

For example, one can take the Standard Model fields
and build all those operators which are invariant under gauge and
Lorentz transformations, including those with a mass dimension larger
than four. This construction is often called the standard model effective
field theory (SMEFT), and it has been studied for a long time. The
unique dimension 5 term in SMEFT, which violates lepton number
and may explain neutrino masses, was mentioned for the first time
in \cite{Weinberg:1979sa}. This very same paper, as well as \cite{Wilczek:1979hc,Abbott:1980zj}
also lists the dimension 6 terms which violate baryon number (hence
they can induce nucleon decay), while the remaining baryon number
conserving operators with this dimension were listed in \cite{Burges:1983zg,Leung:1984ni,Buchmuller:1985jz}.
Nevertheless, for most purposes several of these operators can be
shown to be redundant, and the authors of \cite{Grzadkowski:2010es}
were the first to provide a complete list of non-redundant SMEFT Lagrangian
terms up to dimension 6. The counting of how many parameters are contained
in such terms was given in \cite{Alonso:2013hga}.

It is surprising that it took so long to fully understand these operators,
given that their dimension is not particularly large: most of them
have four fields or less. Currently, the operators of dimension 7
also seem to be well understood \cite{Lehman:2014jma,Liao:2016hru,Liao:2019tep},
but beyond that there are only partial results. For example, some
lepton number violating operators were presented in \cite{Babu:2001ex,deGouvea:2007qla},
and \cite{Fonseca:2018aav} lists field combinations which violate
lepton number in four units but without building the operators explicitly
nor checking for redundancies.

However, in the past few years, there has been a remarkable progress
in the counting of operators, using the Hilbert series \cite{Lehman:2015via,Lehman:2015coa,Henning:2015daa,Henning:2015alf}; see also \cite{Benvenuti:2006qr,Feng:2007ur,Gray:2008yu,Jenkins:2009dy,Hanany:2010vu} for earlier works on this topic.
This technique is not only very elegant and simple to use, but it
can also be applied to an arbitrary model, with an arbitrary 
symmetry group (see for example \cite{Bednyakov:2018cmx,Trautner:2018ipq,Anisha:2019nzx}). Using it, the authors of \cite{Henning:2015alf} computed
the number of SMEFT operators up to dimension 15.

Despite these advantages, the computations performed in the \textit{Hilbert series method} are very
different from those performed in what I will call the \textit{traditional method}, which consists of simply
multiplying all fields in all possible ways, and retaining those combinations
which are invariant under the action of all relevant symmetry groups
--- usually the Lorentz and gauge groups. The choice of the word \textit{traditional} in this context
 is justified by the fact that historically the allowed Lagrangians terms have been found in this way.

There are two complications to this latter approach. One of them is
also present in the Hilbert series method, and it is related to operators
with derivatives: they are problematic mainly because some combinations
of these operators are redundant (under some assumptions) and it would
therefore be desirable to remove them. But it turns out that the ideas
used to handle this issue in Hilbert series computations \cite{Lehman:2015coa,Henning:2015alf}
work equally well when adapted to the traditional method of taking all
invariant fields combinations.

For this reason, the main focus of this document will be the other
complication, which arises when there are repeated fields in an operator.
The simplest example in SMEFT is $LLHH$, which generates neutrino
masses when $H$ acquires a vacuum expectation value. For $n$ copies
of the field $L$, one might have thought that such a term is parameterized
by $n^{2}$ complex numbers. However, this is not true: the coupling
matrix $\kappa$ in flavor space appearing in the expression $\kappa_{ij}L_{i}L_{j}HH$
is symmetric ($\kappa_{ij}=\kappa_{ji}$), so the term is parameterized
by just $n\left(n+1\right)/2$ complex numbers. This is purely a consequence
of the quantum numbers of $L$ and $H$, as I will discuss in detail
later on. It is easy to find more complicated examples, such as
\begin{equation}
N^{c}N^{c}N^{c}N^{c}\textrm{ and }QQQL\,,
\end{equation}
where $N^{c}$ represents right-handed neutrino fields (they do not
exist in SMEFT), which are gauge singlets. While the dimension of
these terms (six) is still quite small, it is non-trivial to derive
the number of independent couplings associated with each, assuming $n_{X}$
flavors of the field type $X=N^{c},Q,L$.\footnote{The answer is $\frac{n_{N^{c}}^{2}}{12}\left(n_{N^{c}}^{2}-1\right)$
and $\frac{n_{Q}n_{L}}{3}\left(2n_{Q}^{2}+1\right)$.}

~

Fortunately, these two complications can be handled in a systematic
way. This makes it possible to count operators in models such as SMEFT,
up to high mass dimensions, using the traditional approach mentioned
earlier. In fact, besides the number of operators, it is also possible
to extract the symmetry of coupling tensors (such as $\kappa$ above)
under flavor index permutations; in some cases, this is a piece of
extra information which cannot be inferred from the number of operators.

It is also noteworthy that the traditional method described in this
work seems to be computationally competitive with the Hilbert series
approach. The authors of \cite{Henning:2015alf} provided the number
of each type of SMEFT operator up to dimension 12 plus the total number
of operators with dimensions $d=13,14$ and $15$. Up to now, these
results had only been cross-checked by other means up to $d=8$. However,
with the method explained in this work it was possible to count all
types of SMEFT operators up to dimension 15 in a couple of hours,
using a standard laptop (all numbers given in \cite{Henning:2015alf}
were reproduced). Existing Mathematica code \cite{Fonseca:2011sy,Fonseca:2017lem}
was modified to make such calculations possible for any model, and
such code is publicly available at the web address
\begin{center}
\texttt{renatofonseca.net/sym2int.php}
\par\end{center}

The remainder of the text is structured as follows:
\begin{itemize}
\item Section \ref{sec:2} describes the notation and the conventions adopted
in this work.
\item Section \ref{sec:3} is devoted to the problem of repeated fields
in interactions, suggesting a way of systematically dealing with these
cases.
\item Section \ref{sec:4-derivatives} discusses the problem with derivatives,
and how they can be handled by using and adapting the solutions proposed
in \cite{Lehman:2015coa,Henning:2015alf}.
\item Section \ref{sec:5} contains a discussion of several topics related
to the counting of operators of an effective field theory. The ideas
mentioned in sections 3 and 4 were implemented in a computer code,
and some of the results obtained with it are presented here and in
an appendix.
\item A summary is available at the end.
\end{itemize}

\section{\label{sec:2}Notation and conventions}

The method discussed in this work can be applied to any effective
field theory, nevertheless SMEFT will often serve as an example. Its
Lagrangian is invariant under the $SU(3)_{C}\times SU(2)_{L}\times U(1)_{Y}$
gauge group and the restricted Lorentz group\footnote{The word \textit{restricted} here refers to the fact that time-reversal
($T$) and space-inversion ($P$) transformations are not included.
This group is usually denoted by $SO^{+}\left(1,3\right)$.}. For all practical purposes the algebra of this latter group can
be taken to be the same as the one of $SU(2)_{l}\times SU(2)_{r}$
so its representations are given by pairs of non-negative half-integers
$\left(j_{l},j_{r}\right)$. The only caveat to keep in mind is
that complex conjugation flips these numbers: $\left(j_{l},j_{r}\right)^{*}=\left(j_{r},j_{l}\right)$.
In particular, the representations $\left(0,0\right)$, $\left(\frac{1}{2},0\right)$,
$\left(0,\frac{1}{2}\right)$ and $\left(\frac{1}{2},\frac{1}{2}\right)$
correspond to a scalar, a left-handed Weyl spinor, a right-handed
Weyl spinor and a 4-vector, in this order. A field strength tensor
$\mathcal{F}_{\mu\nu}$ transforms under the Lorentz group in the
same way as the anti-symmetric part of the product of two 4-vectors,
i.e.
\begin{equation}
\left[\left(\frac{1}{2},\frac{1}{2}\right)\times\left(\frac{1}{2},\frac{1}{2}\right)\right]_{A}=\left(1,0\right)+\left(0,1\right)\,.
\end{equation}
Note that $\mathcal{F}_{\mu\nu}$ is a real field so the three components
transforming as $\left(0,1\right)$ are the complex conjugate of those
transforming as $\left(1,0\right)$. I will refer to the part transforming
as $\left(1,0\right)$ simply with the letter $F$ and the other part
as $F^{*}$ (so $\mathcal{F}_{\mu\nu}=F+F^{*}$). A subscript might
be added to $F$ to identify with which of the gauge factor groups it
is associated.

All SMEFT operators are combinations of the fields shown in table
\ref{tab:SMEFT-field-content}, their conjugates and their derivatives
(which need to be covariant). There are 3 copies (or flavors) of each
fermion field, but we may just treat this number as a variable $n$.

\begin{table}[h]
\begin{centering}
\begin{tabular}{cccccc}
\toprule 
 & $SU(3)_{C}$ & $SU(2)_{L}$ & $U(1)_{Y}$ & $SU(2)_{l}\times SU(2)_{r}$ & \# flavors\tabularnewline
\midrule
$Q$ & $\boldsymbol{3}$ & $\boldsymbol{2}$ & $\frac{1}{6}$ & $\left(\frac{1}{2},0\right)$ & $n$\tabularnewline
$u^{c}$ & $\overline{\boldsymbol{3}}$ & $\boldsymbol{1}$ & $-\frac{2}{3}$ & $\left(\frac{1}{2},0\right)$ & $n$\tabularnewline
$d^{c}$ & $\overline{\boldsymbol{3}}$ & $\boldsymbol{1}$ & $\frac{1}{3}$ & $\left(\frac{1}{2},0\right)$ & $n$\tabularnewline
$L$ & $\boldsymbol{1}$ & $\boldsymbol{2}$ & $-\frac{1}{2}$ & $\left(\frac{1}{2},0\right)$ & $n$\tabularnewline
$e^{c}$ & $\boldsymbol{1}$ & $\boldsymbol{1}$ & $1$ & $\left(\frac{1}{2},0\right)$ & $n$\tabularnewline
$H$ & $\boldsymbol{1}$ & $\boldsymbol{2}$ & $\frac{1}{2}$ & $\left(0,0\right)$ & $1$\tabularnewline
$F_{G}$ & $\boldsymbol{8}$ & $\boldsymbol{1}$ & 0 & $\left(1,0\right)$ & $1$\tabularnewline
$F_{W}$ & $\boldsymbol{1}$ & $\boldsymbol{3}$ & 0 & $\left(1,0\right)$ & $1$\tabularnewline
$F_{B}$ & $\boldsymbol{1}$ & $\boldsymbol{1}$ & 0 & $\left(1,0\right)$ & $1$\tabularnewline
\bottomrule
\end{tabular}
\par\end{centering}
\caption{\label{tab:SMEFT-field-content}SMEFT field content. All operators
are built from them, their conjugates and their derivatives. The number
of fermion flavors $n$, which is 3, will often be left unspecified.}

\end{table}

Finally, concerning conventions, it is worth noting that sometimes
the word \textit{\small{}operator} is used in reference to different
things:
\begin{enumerate}
\item It might refer to the different gauge and Lorentz invariant contraction
of fields, with the flavour indices expanded. In that case the lepton
Yukawa interactions $L_{i}e_{j}^{c}H^{*}$ correspond to 9 complex
operators, or 18 real ones. That is the meaning I will use for the
word \textit{\small{}operator} (which might be real or complex). With
this understanding, there are 3045 dimension 6 real operators in SMEFT
(546 of which violate baryon number) \cite{Alonso:2013hga,Henning:2015alf}.
\item An alternative view is to see some groups of {\small{}operators} such
as the lepton Yukawa interactions $L_{i}e_{j}^{c}H^{*}$ as a single
structure which I will call a (Lagrangian) \textit{\small{}term};
they can be thought of as {\small{}operators} with the flavor indices
unexpanded. With this terminology, SMEFT can be written with just
84 real terms of mass dimension 6 \cite{Grzadkowski:2010es}.
\item The most general interaction of some combinations of fields cannot
be written in the Lagrangian as a single term. In other words, a single
coupling tensor $\kappa{}_{ij\cdots}$ with indices $ij\cdots$ contracted
with the field flavors is not enough. That is what happens, for example,
with the fields $Q^{*}L^{*}QL$: a minimum of two term $\kappa^{(\alpha)}{}_{ijkl}\left(Q_{i}^{*}L_{j}^{*}Q_{k}L_{l}\right)_{(\alpha)}$,
$\alpha=1,2$ are needed. We may refer to each collection of terms
associated with a common product of fields as a \textit{type of operator}.\footnote{So each operator type is defined by the number of occurrences $m_{i}$
of the various kinds of field $\chi_{i}$ in the model ($=Q,u^{c},d^{c}\dots$
in SMEFT). For these purposes, a derivative may be considered as a
field, so we should just count them.} There are 72 real types of operators with dimension 6 in SMEFT.
\end{enumerate}

\section{\label{sec:3}Operators with repeated fields}

\subsection{The problem}

If there are no derivatives nor repeated fields, the counting of operators
of a certain \textit{type} $\chi^{(1)}\chi^{(2)}\cdots\chi^{(m)}$
is straightforward:
\begin{itemize}
\item Based on the quantum numbers of the fields, we may calculate the number
$t$ of independent gauge and Lorentz invariant contractions of their
components. Each yields one \textit{term}. These contractions can be found systematically with the method of weights  \cite{Slansky:1981yr,Cahn:1985wk} or, in the case of special unitary groups, with the tensor method  (see for instance chapter 4 of \cite{Cheng:1985bj}).
\item Each term is associated with $\prod_{i=1}^{m}n_{i}$ \textit{operators},
where $n_{i}$ is the number of flavors of the field $\chi^{(i)}$.
\item The number quoted above corresponds to real operators if conjugating
the set of fields $\left\{ \chi^{(i)}\right\} $ yields back this
very same set, i.e.$\left\{ \chi^{(i)}\right\} ^{*}=\left\{ \chi^{(i)}\right\} $,
otherwise there are $\prod_{i=1}^{m}n_{i}$ complex operators in each
term (which is the same as saying that there are $2\prod_{i=1}^{m}n_{i}$
real operators).
\end{itemize}
So, in the absence of derivatives and repeated fields, there are $c\,t\prod n_{i}$
real operators of type $\chi^{(1)}\chi^{(2)}\cdots\chi^{(m)}$, where
$c=1$ if $\left\{ \chi^{(i)}\right\} ^{*}=\left\{ \chi^{(i)}\right\} $
and $c=2$ otherwise. For example, there are $2n^{4}=162$ (for $n=3$)
real operators of type $Q^{*}L^{*}QL$.\footnote{The list of fields is self-conjugate, $\left\{ Q^{*},L^{*},Q,L\right\} ^{*}=\left\{ Q^{*},L^{*},Q,L\right\} $,
so $c=1$. On the other hand, there are 2 ways of making the $SU(2)_{L}$
contractions of the fields, so $t=2$.}

~

The above is straightforward to compute. Nevertheless, repeated fields
complicate this analysis significantly (the problem with derivatives
will be discussed in the next section).

Let us start by taking a close look at the type of operator $LLHH$,
which is not associated with $n^{2}$ complex operators, but rather
with $n\left(n+1\right)/2$ as pointed out earlier. Given that $L$
and $H$ are both doublets of $SU(2)_{L}$, there are two possible
contractions: $\left(LL\right)_{\boldsymbol{1}}\left(HH\right)_{\boldsymbol{1}}$
and $\left(LL\right)_{\boldsymbol{3}}\left(HH\right)_{\boldsymbol{3}}$.
In one of them, the two $L$'s are contracted together as a singlet
of $SU(2)_{L}$, and the same happens with the Higgs fields. The other
possibility is for the $L$'s and $H$'s to contract as triplets.
Crucially, the singlet contraction of two doublets is anti-symmetric,
while the triplet contraction is symmetric. Since there is only one
copy of the Higgs field, we retain only the latter. If we now introduce
flavor indices and a parameter tensor $\kappa$ we obtain the term
\begin{equation}
\kappa_{ij}\left(L_{i}L_{j}\right)_{\boldsymbol{3}}\left(HH\right)_{\boldsymbol{3}}\,.
\end{equation}
This is the same as $\kappa_{ij}\left(L_{j}L_{i}\right)_{\boldsymbol{3}}\left(HH\right)_{\boldsymbol{3}}$
so, renaming the dummy indices $i$ and $j$, we can also write it
as $\kappa_{ji}\left(L_{i}L_{j}\right)_{\boldsymbol{3}}\left(HH\right)_{\boldsymbol{3}}$.
We then conclude that only the symmetric part of $\kappa$ is relevant:\footnote{One could add an anti-symmetric part to this matrix $\kappa$, but
it would not affect the Lagrangian. So it is pointless to do so.}
\begin{equation}
\kappa_{ij}=\kappa_{ji}\,.
\end{equation}
Such a matrix has $n\left(n+1\right)/2$ complex degrees of freedom.

Now let us go beyond SMEFT for a moment in order to consider what
would happen to the type of operator $LLHH$ if there were multiple
copies of the Higgs field. For that purpose we may add flavor indices
to $H$, noting that now the contraction $\left(LL\right)_{\boldsymbol{1}}\left(HH\right)_{\boldsymbol{1}}$
is not null so we can have two terms:
\begin{equation}
\kappa_{ijkl}^{(1)}\left(L_{i}L_{j}\right)_{\boldsymbol{1}}\left(H_k H_l\right)_{\boldsymbol{1}}+\kappa_{ijkl}^{(3)}\left(L_{i}L_{j}\right)_{\boldsymbol{3}}\left(H_k H_l\right)_{\boldsymbol{3}}\,.\label{eq:4}
\end{equation}
It should be clear from the discussion so far that the tensors $\kappa^{(1)}$
and $\kappa^{(2)}$ have symmetries: 
\begin{align}
\kappa_{ijkl}^{(1)} & =-\kappa_{jikl}^{(1)}=-\kappa_{ijlk}^{(1)}\,,\\
\kappa_{ijkl}^{(3)} & =\kappa_{jikl}^{(3)}=\kappa_{ijlk}^{(3)}\,.
\end{align}
As a consequence, $\kappa^{(1)}$ and $\kappa^{(3)}$ contain
\begin{equation}
\left[n\left(n-1\right)/2\right]\left[n_{H}\left(n_{H}-1\right)/2\right]
\end{equation}
and
\begin{equation}
\left[n\left(n+1\right)/2\right]\left[n_{H}\left(n_{H}+1\right)/2\right]
\end{equation}
complex parameters, respectively ($n_{H}$ being the number of Higgs
copies). But do we need two terms as in expression (\ref{eq:4})?
The answer is no; there is no such need. Take a single term $\mathcal{O}$
formed from a linear combination of $\mathcal{O}^{(1)}\equiv\left(LL\right)_{\boldsymbol{1}}\left(HH\right)_{\boldsymbol{1}}$
and $\mathcal{O}^{(3)}\equiv\left(LL\right)_{\boldsymbol{3}}\left(HH\right)_{\boldsymbol{3}}$:
\begin{equation}
\mathcal{O}=\alpha_{1}\mathcal{O}^{(1)}+\alpha_{3}\mathcal{O}^{(3)}\quad\left(\alpha_{1},\alpha_{3}\neq0\right)\,.
\end{equation}
We can recover $\mathcal{O}^{(1)}$ and $\mathcal{O}^{(3)}$
from $\mathcal{O}$ by (anti)symmetrizing the flavor indices, $\mathcal{O}_{ijkl}^{(1)}=\alpha_{1}^{-1}\mathcal{O}_{\left[ij\right]\left[kl\right]}$
and $\mathcal{O}_{ijkl}^{(3)}=\alpha_{3}^{-1}\mathcal{O}_{\left(ij\right)\left(kl\right)}$,
and therefore we can write 
\begin{equation}
\kappa_{ijkl}\mathcal{O}_{ijkl}
\end{equation}
instead of expression (\ref{eq:4}). Nevertheless, the tensor $\kappa$
is not fully generic, as it contains only $nn_{H}\left(nn_{H}+1\right)/2$
independent complex parameters (the sum of the number of parameters
in $\kappa^{(1)}$ and $\kappa^{(3)}$). In fact, we may write
\begin{equation}
\kappa=\alpha_{1}^{-1}\kappa^{(1)}+\alpha_{3}^{-1}\kappa^{(3)}\,
\end{equation}
or, alternatively, $\kappa$ can be described as the most general
tensor with the symmetry $\kappa_{ijkl}=\kappa_{jilk}$.

Returning to SMEFT, an identical situation happens with the type of
operator $L^{*}L^{*}LL$: four $SU(2)_{L}$ doublets can be contracted
in two different ways \cite{Buchmuller:1985jz}, but only one term
is required \cite{Grzadkowski:2010es}, and it is associated with a
total of $n^{2}\left(n^{2}+1\right)/2=45$ complex operators \cite{Alonso:2013hga}
(this last expression is the same as the one presented before in relation
to $LLHH$, for the special case where $n_{H}=n$).

These examples highlight the following points:
\begin{itemize}
\item The symmetry of the coupling tensors $\kappa$ is determined by the
quantum numbers of the fields in each term.
\item From the symmetry of the coupling tensors we may derive the number
of independent operators.
\item Unlike the number of independent operators, the number of terms is
an ambiguous quantity since it is possible to merge multiple terms
into a single one. To get around this ambiguity, we may always focus
on writing a Lagrangian with a minimal number of terms.
\item Merging terms might seem convenient, but there is a drawback to doing
so: it may become more complicated to understand the structure of
the associated coupling tensor $\kappa$.
\end{itemize}
In order to derive the symmetry of the coupling tensors under permutations
of indices, we may consult tables such as in \cite{Slansky:1981yr} which
list the permutation symmetry of products of representations. For
example, in the case of $SU(2)$ it is well known that
\begin{equation}
\boldsymbol{2}\times\boldsymbol{2}=\boldsymbol{1}_{A}+\boldsymbol{3}_{S}\,,\label{eq:10}
\end{equation}
with the subscripts indicating that each component in the product
is either symmetric (S) or anti-symmetric (A) under a switch of the
two doublets.\footnote{To be explicit, it is well known that if $\left(2_{1},2_{2}\right)^{T}$
and $\left(2_{1}^{\prime},2_{2}^{\prime}\right)^{T}$ are two doublets,
then the combination $2_{1}2_{2}^{\prime}-2_{2}2_{1}^{\prime}$ is
$SU(2)$ invariant, while $\left(2_{1}2_{1}^{\prime},\frac{1}{\sqrt{2}}2_{1}2_{2}^{\prime}+\frac{1}{\sqrt{2}}2_{2}2_{1}^{\prime},2_{2}2_{2}^{\prime}\right)^{T}$
transforms as a triplet. Switching the two doublets flips the sign
of the first expression, while the triplet remains unchanged.} However, such tables might not be enough: consider for example the
product of four scalar $SU(2)$ doublets. There are two invariant
contractions because the product of four doublets contains two singlets:
\begin{equation}
\boldsymbol{2}\times\boldsymbol{2}\times\boldsymbol{2}\times\boldsymbol{2}=\boldsymbol{1}+\boldsymbol{1}+\cdots\,.\label{eq:11}
\end{equation}
Are these singlets symmetric (S) or antisymmetric (A)? Mathematically,
the answer is clear but it might be somewhat confusing if one is unaware
of this type of complications: the singlets are neither (completely)
symmetric nor anti-symmetric. Let us go back to equation (\ref{eq:10});
as far as subscripts are concerned, all we need to know is what happens
to each summand in the right-hand side under the permutation of the
two doublets. The permutation of $m$ objects forms a discrete group
of size $m!$ usually denoted by $S_{m}$, and in the particular case
of $S_{2}$ everything is very simple. There are two elements in the
group, the identity $e$ (do nothing) and $a$ (transpose the two
objects), with $a^{2}=e$. There are also only two irreducible representations,
which are both 1-dimensional: the trivial/symmetric one (S) under
which $x\mapsto x$, and the alternating/anti-symmetric representation
(A) under which $x\mapsto\left(-1\right)^{\pi}x$, where $\left(-1\right)^{\pi}=+1$
if $\pi=e$ and $=-1$ if $\pi=a$. In other words, the action of
the group $S_{2}$ can always be translated into signs $\pm1$: in
equation (\ref{eq:10}) the singlet $\boldsymbol{1}$ transforms under
$S_{2}$ as the alternating representation, while the components of
the triplet $\boldsymbol{3}$ transform trivially.

In the case of equation (\ref{eq:11}) we must find out what happens
to its right-hand side under arbitrary permutations of the four doublets,
so we must deal with the irreducible representations of the $S_{4}$
group. It turns out that the singlets in equation (\ref{eq:11}) transform
as an irreducible 2-dimensional representation of $S_{4}$. To better
grasp the significance of this statement, let us see these two $SU(2)$-invariant
contractions of four doublets explicitly. We can write them as follows:
\begin{align}
c^{(1)} & =\epsilon_{ij}\epsilon_{kl}2_{i}2_{j}^{\prime}2_{k}^{\prime\prime}2_{l}^{\prime\prime\prime}\,,\label{eq:12}\\
c^{(2)} & =\epsilon_{ik}\epsilon_{jl}2_{i}2_{j}^{\prime}2_{k}^{\prime\prime}2_{l}^{\prime\prime\prime}\,,\label{eq:13}
\end{align}
where the subscripts take the value 1 or 2 (these are doublet indices
unrelated to flavor) and $\epsilon$ is the Levi-Civita tensor. Note
than even though there are $m!$ permutations which can be performed
on $m$ objects $x_{i}$, all of them can be obtained through successive
applications of only two of them: $x_{1}\rightarrow x_{2}\rightarrow x_{1}$
and $x_{1}\rightarrow x_{2}\rightarrow x_{3}\rightarrow\cdots\rightarrow x_{m}\rightarrow x_{1}$.
So, let us consider in the following what happens to $c^{(1)}$ and
$c^{(2)}$ under the changes \textbf{$\boldsymbol{2}\leftrightarrow\boldsymbol{2^{\prime}}$}
and $\boldsymbol{2}\rightarrow\boldsymbol{2^{\prime}}\rightarrow\boldsymbol{2^{\prime\prime}}\rightarrow\boldsymbol{2^{\prime\prime\prime}}\rightarrow\boldsymbol{2}$.
It is rather easy to see that
\begin{align}
\left(\begin{array}{c}
c^{(1)}\\
c^{(2)}
\end{array}\right)_{\boldsymbol{2}\leftrightarrow\boldsymbol{2^{\prime}}} & =\left(\begin{array}{cc}
-1 & 0\\
-1 & 1
\end{array}\right)\cdot\left(\begin{array}{c}
c^{(1)}\\
c^{(2)}
\end{array}\right)\,,\label{eq:14}\\
\left(\begin{array}{c}
c^{(1)}\\
c^{(2)}
\end{array}\right)_{\!\!\!\!\!\begin{array}{c}
{\scriptstyle \boldsymbol{2}\rightarrow\boldsymbol{2^{\prime}}\rightarrow}\\
{\scriptstyle \boldsymbol{2^{\prime\prime}}\rightarrow\boldsymbol{2^{\prime\prime\prime}}\rightarrow\boldsymbol{2}}
\end{array}} & =\left(\begin{array}{cc}
1 & -1\\
0 & -1
\end{array}\right)\cdot\left(\begin{array}{c}
c^{(1)}\\
c^{(2)}
\end{array}\right)\,.\label{eq:15}
\end{align}
No change of basis $c_{new}^{(\alpha)}=B_{\alpha\beta}c^{(\beta)}$
will simultaneously diagonalize the two matrices above, so $c^{(1),(2)}$
form an irreducible 2-dimensional representation of the permutation
group $S_{4}$. As a consequence, in this case the effect of permuting
the doublets cannot be reduced to a simple matter of signs $\pm1$.

We now introduce flavor, so there will be two sets of indices: the
$g_{i}$ will stand for group indices, while flavor indices will be
called $f_{i}$. For scalar doublets $\phi$ with multiple flavors
we should write down two terms:

\begin{equation}
\left(\kappa_{f_{1}f_{2}f_{3}f_{4}}^{(1)}c_{g_{1}g_{2}g_{3}g_{4}}^{(1)}+\kappa_{f_{1}f_{2}f_{3}f_{4}}^{(2)}c_{g_{1}g_{2}g_{3}g_{4}}^{(2)}\right)\phi_{g_{1}}^{f_{1}}\phi_{g_{2}}^{f_{2}}\phi_{g_{3}}^{f_{3}}\phi_{g_{4}}^{f_{4}}\,,\label{eq:4-doublets}
\end{equation}
where $c_{g_{1}g_{2}g_{3}g_{4}}^{(1),(2)}$ represent the tensors
in front of the doublets in equations (\ref{eq:12}) and (\ref{eq:13}).
As for $\kappa^{(1)}$ and $\kappa^{(2)}$, they are tensors containing
free parameters, and there is some symmetry associated with them, which
is yet to be determined.

Instead of writing down all the indices in equation (\ref{eq:4-doublets})
explicitly, we may use the short-hand notation
\begin{equation}
\kappa_{\left\{ f\right\} }^{(\alpha)}c_{\left\{ g\right\} }^{(\alpha)}\phi_{\left\{ g\right\} }^{\left\{ f\right\} }\,,\label{eq:17}
\end{equation}
where repeated indices are summed over. We know two things:
\begin{enumerate}
\item $\phi_{\left\{ g\right\} }^{\left\{ f\right\} }$ is symmetric under
equal permutation of the $f$ and $g$ indices, i.e. $\phi_{\pi\left\{ g\right\} }^{\pi\left\{ f\right\} }=\phi_{\left\{ g\right\} }^{\left\{ f\right\} }$
for any permutation $\pi$;
\item $c_{\pi\left\{ g\right\} }^{(\alpha)}=P\left(\pi^{-1}\right)_{\alpha\beta}c_{\left\{ g\right\} }^{(\beta)}$,
where the two-by-two matrices $P\left(\pi^{-1}\right)$ can be obtained,
for any $\pi$, from the product of the two matrices in equations
(\ref{eq:14}) and (\ref{eq:15}).\footnote{The inverse permutation of $\pi$, $\pi^{-1}$ ($\pi^{-1}\circ\pi=id$),
appears here rather than $\pi$ itself for the following reason, which
is not important for the present discussion. Strictly speaking, for
each $\alpha$ the numbers $c_{\left\{ g\right\} }^{(\alpha)}$ are
not a tensor but rather the components of a tensor in some basis $\left|e_{g_{1}}e_{g_{2}}\cdots e_{g_{m}}\right\rangle \equiv\left|e_{\left\{ g\right\} }\right\rangle $.
In other words, the tensors are $C^{(\alpha)}=c_{\left\{ g\right\} }^{(\alpha)}\left|e_{\left\{ g\right\} }\right\rangle $.
Under a permutation $\pi$, the basis of the tensors changes, $\left|e_{\left\{ g\right\} }\right\rangle \rightarrow\pi\left(\left|e_{\left\{ g\right\} }\right\rangle \right)=\left|e_{\pi\left\{ g\right\} }\right\rangle $,
so $\pi\left(C^{(\alpha)}\right)=c_{\left\{ g\right\} }^{(\alpha)}\left|e_{\pi\left\{ g\right\} }\right\rangle =c_{\pi^{-1}\left\{ g\right\} }^{(\alpha)}\left|e_{\left\{ g\right\} }\right\rangle $.
However, we know that this result must be a linear combination of
the $C^{(\alpha)}$, $P\left(\pi\right)_{\alpha\beta}C^{(\beta)}$,
and therefore we conclude that $c_{\pi\left\{ g\right\} }^{(\alpha)}=P\left(\pi^{-1}\right)_{\alpha\beta}c_{\left\{ g\right\} }^{(\beta)}$.}
\end{enumerate}
It follows that
\begin{equation}
\kappa_{\left\{ f\right\} }^{(\alpha)}c_{\left\{ g\right\} }^{(\alpha)}\phi_{\left\{ g\right\} }^{\left\{ f\right\} }=\kappa_{\pi\left\{ f\right\} }^{(\alpha)}c_{\pi\left\{ g\right\} }^{(\alpha)}\phi_{\pi\left\{ g\right\} }^{\pi\left\{ f\right\} }=\kappa_{\pi\left\{ f\right\} }^{(\alpha)}P\left(\pi^{-1}\right)_{\alpha\beta}c_{\left\{ g\right\} }^{(\beta)}\phi_{\left\{ g\right\} }^{\left\{ f\right\} }\,,
\end{equation}
so
\begin{equation}
\kappa_{\pi\left\{ f\right\} }^{(\alpha)}P\left(\pi^{-1}\right)_{\alpha\beta}=\kappa_{\left\{ f\right\} }^{(\beta)}\textrm{ or equivalently }\kappa_{\pi\left\{ f\right\} }^{(\alpha)}=\left[P\left(\pi\right)^{T}\right]_{\alpha\beta}\kappa_{\left\{ f\right\} }^{(\beta)}\,.
\end{equation}
To be explicit,
\begin{align}
\left(\begin{array}{c}
\kappa_{f_{2}f_{1}f_{3}f_{4}}^{(1)}\\
\kappa_{f_{2}f_{1}f_{3}f_{4}}^{(2)}
\end{array}\right) & =\left(\begin{array}{cc}
-1 & -1\\
0 & 1
\end{array}\right)\cdot\left(\begin{array}{c}
\kappa_{f_{1}f_{2}f_{3}f_{4}}^{(1)}\\
\kappa_{f_{1}f_{2}f_{3}f_{4}}^{(2)}
\end{array}\right)\,,\\
\left(\begin{array}{c}
\kappa_{f_{2}f_{3}f_{4}f_{1}}^{(1)}\\
\kappa_{f_{2}f_{3}f_{4}f_{1}}^{(2)}
\end{array}\right) & =\left(\begin{array}{cc}
1 & 0\\
-1 & -1
\end{array}\right)\cdot\left(\begin{array}{c}
\kappa_{f_{1}f_{2}f_{3}f_{4}}^{(1)}\\
\kappa_{f_{1}f_{2}f_{3}f_{4}}^{(2)}
\end{array}\right)\,.
\end{align}
These are the only constraints on the parameter tensors $\kappa^{(1)}$
and $\kappa^{(2)}$, so these objects can be seen as the most general
tensors which satisfy the above equations. It is not very useful for
our purposes to find their exact form, which is basis dependent in
any case ($\kappa^{(1)}$ and $\kappa^{(2)}$ would change under a
redefinition $c_{new}^{(\alpha)}=B_{\alpha\beta}c^{(\beta)}$). However,
just as with the $LLHH$ type of operator, it would be interesting
to know how many operators there are in expression (\ref{eq:4-doublets})
or, in other words, how many independent entries there are in the
$\kappa^{(\alpha)}$ tensors. The answer, which is $n_{\phi}^{2}\left(n_{\phi}^{2}-1\right)/12$
for $n_{\phi}$ flavors of $\phi$, depends only on the fact that
the $c^{(\alpha)}$ (and consequently the $\kappa^{(\alpha)}$ as
well) transform under the 2-dimensional representation of $S_{4}$.
We shall discuss below how to compute these numbers.

There is one last remark to be made about this example. It might have
seemed that we need to write down two terms as shown in expression
(\ref{eq:4-doublets}), but it is easy to see that they can be merged
into a single one. That is because we may write one of the $c^{(\alpha)}$
as a function of the other, with the indices permuted. In particular,
from equations (\ref{eq:14}) and (\ref{eq:15}) it can be inferred
that $c_{g_{1}g_{2}g_{3}g_{4}}^{(2)}=c_{g_{1}g_{3}g_{2}g_{4}}^{(1)}$
or, in a short-hand notation, $c_{\pi_{23}\left\{ g\right\} }^{(2)}=c_{\left\{ g\right\} }^{(1)}$.
Using the fact that $\phi_{\pi\left\{ g\right\} }^{\pi\left\{ f\right\} }=\phi_{\left\{ g\right\} }^{\left\{ f\right\} }$,
we obtain that
\begin{equation}
\left(\kappa_{\left\{ f\right\} }^{(1)}c_{\left\{ g\right\} }^{(1)}+\kappa_{\left\{ f\right\} }^{(2)}c_{\left\{ g\right\} }^{(2)}\right)\phi_{\left\{ g\right\} }^{\left\{ f\right\} }=\left(\kappa_{\left\{ f\right\} }^{(1)}+\kappa_{\pi_{23}\left\{ f\right\} }^{(2)}\right)c_{\left\{ g\right\} }^{(1)}\phi_{\left\{ g\right\} }^{\left\{ f\right\} }\equiv\kappa_{\left\{ f\right\} }^{\textrm{merged}}c_{\left\{ g\right\} }^{(1)}\phi_{\left\{ g\right\} }^{\left\{ f\right\} }\,.
\end{equation}
Note that $\kappa^{\textrm{merged}}$ is not a completely generic
4-index tensor since it has only $n_{\phi}^{2}\left(n_{\phi}^{2}-1\right)/12$
independent entries, rather than $n_{\phi}^{4}$.

\subsection{The permutation group of $m$ objects}

These examples highlight the fact that in order to list operators
and terms in the Lagrangian, it is important to have an understanding
of the permutation group $S_{m}$. As such, I will briefly review
here some of its properties.

The irreducible representations of $S_{m}$ can be labeled by the
different ways in which the number $m$ can be partitioned. For example
\begin{equation}
4=3+1=2+2=2+1+1=1+1+1+1\,,
\end{equation}
so there are five partitions $\left\{ 4\right\} $, $\left\{ 3,1\right\} $,
$\left\{ 2,2\right\} $, $\left\{ 2,1,1\right\} $ and $\left\{ 1,1,1,1\right\} $
and these can be identified with the five irreducible representations
of $S_{4}$. It is common to depict the partition $\lambda=\left\{ \lambda_{1},\lambda_{2},\cdots\right\} $
of $m$ (symbolically $\lambda\vdash m$) with $\lambda_{1}$ boxes
in a row, followed by $\lambda_{2}$ boxes in a second row, and so
on, such that $\sum_{i}\lambda_{i}=m$ and $\lambda_{i+1}\leq\lambda_{i}$.
These are called Young diagrams. For example:\ytableausetup{centertableaux}\ytableausetup{boxsize=0.7em}
\begin{gather}
\left\{ 4\right\} =\ydiagram{4}\quad\left\{ 3,1\right\} =\ydiagram{3,1}\quad\left\{ 2,2\right\} =\ydiagram{2,2}\quad\left\{ 2,1,1\right\} =\ydiagram{2,1,1}\quad\left\{ 1,1,1,1\right\} =\ydiagram{1,1,1,1}\,.
\end{gather}
The product of $S_{m}$ representations can easily and efficiently
be decomposed in irreducible parts as in this example:
\begin{gather}
\left(\;\ydiagram{2,2}\;\right)^{3}\times\left(\;\ydiagram{3,1}\;\right)^{5}=80\;\ydiagram{4}+244\;\ydiagram{3,1}\;+160\;\ydiagram{2,2}+244\;\ydiagram{2,1,1}+80\;\ydiagram{1,1,1,1}\,.
\end{gather}
To perform such decomposition, for $S_n$  or any other finite group, it suffices to know the group's character table.\footnote{Readers
	 interested in knowing more about this are referred to group theory textbooks (for instance \cite{Tung-book}). It is worth noting that specific techniques can be applied to the $S_n$ family of groups (see for example \cite{Bernstein:2003}). In any case, one can also use pre-calculated character tables or readily available computer codes, such as \texttt{GroupMath}'s function \texttt{SnClassCharacter[lambda,mu]} which calculates the character of the class \texttt{mu} in the irreducible representation \texttt{lambda}.}
 
There are two other important properties of the irreducible representations
$\lambda$ of $S_{m}$ which are worth mentioning. First, the dimension
of $\lambda$ can be calculated with the famous Hook length formula \cite{Hook-length-formula-1954},
\begin{equation}
d\left(\lambda\right)=\frac{m!}{\prod_{u}h\left(u\right)}\,,
\end{equation}
where $u$ represents each box of the Young diagram of $\lambda$
and $h\left(u\right)$ is equal to the number of boxes to the right
of $u$ plus the number of boxes below $u$ plus 1:\ytableausetup{boxsize=1.5em}\begin{gather}h(u):\;\begin{ytableau}4 & 2 & 1 \\ 1\end{ytableau}\Rightarrow d\left(\left\{ 3,1\right\} \right)=\frac{4!}{4.2.1.1}=3\,.\end{gather}

The second important property of a partition $\lambda$ is the following.
A semi-standard Young tableaux with shape $\lambda$ is obtained by
filling the Young diagram of $\lambda$ with natural numbers up to
some limit $n$ (possibly repeating these numbers) in such a way that
the numbers increase along each column, and they do not decrease along
rows. For example, in the case of $\lambda=\left\{ 2,2\right\} $
and $n=3$ there are six possibilities:\begin{gather}\begin{ytableau}1 & 1 \\ 2&2\end{ytableau}\,,\;\begin{ytableau}1 & 1 \\ 2&3\end{ytableau}\,,\;\begin{ytableau}1 & 1 \\ 3&3\end{ytableau}\,,\;\begin{ytableau}1 & 2 \\ 2&3\end{ytableau}\,,\;\begin{ytableau}1 & 2 \\ 3&3\end{ytableau}\,,\;\begin{ytableau}2 & 2 \\ 3&3\end{ytableau}\,\,.\end{gather}The
number of semi-standard Young tableaux with a shape $\lambda$ using
numbers up to $n$ can be calculated with the Hook content formula
(see \cite{Stanley:1999}),
\begin{equation}
\mathcal{S}\left(\lambda,n\right)=\prod_{u}\frac{n+c\left(u\right)}{h\left(u\right)}\,,\label{eq:S}
\end{equation}
where $u$ represent each box of the diagram, $h\left(u\right)$ was
mentioned earlier, and $c\left(u\right)$ is equal to the column number
minus the row number of $u$:\begin{gather}c(u):\;\begin{ytableau}0 & 1 & 2 \\ -1\end{ytableau}\,.\end{gather}As
an example, notice that $\left\{ 2,2\right\} $ is the 2-dimensional
representation of $S_{4}$, $d\left(\left\{ 2,2\right\} \right)=2$,
and furthermore $\mathcal{S}\left(\left\{ 2,2\right\} ,n_{\phi}\right)=n_{\phi}^{2}\left(n_{\phi}^{2}-1\right)/12$
using formula (\ref{eq:S}). As we have seen, this last expression
corresponds to the number of quartic operators which can be build
with $n_{\phi}$ scalars $\phi$ transforming as doublets of $SU(2)$.

\subsection{Permuting the indices of tensors}

Consider now a tensor $T_{i_{1}i_{2}\cdots i_{m}}$ where each index
goes from 1 to some numbers $n$. It is well known to mathematicians
that the $n^{m}$ components can either be referenced with the $m$
indices $\left(i_{1},i_{2},\cdots,i_{m}\right)$ or alternative by
just 3 special indices $\left(\lambda,\alpha,j\right)$,
\begin{equation}
T_{i_{1}i_{2}\cdots i_{m}}\leftrightarrow T_{\lambda\alpha j}\,.
\end{equation}
Rather than trying to write explicitly the components of $T_{\lambda\alpha j}$
as a function of those of $T_{i_{1}i_{2}\cdots i_{m}}$, it is more
illuminating to describe the nature of the 3 new labels. The first
one ($\lambda$) can be any partition of $m$, the second one ($\alpha$)
takes values from 1 to $\mathcal{S}\left(\lambda,n\right)$, and the
last one ($j$) goes from 1 to $d\left(\lambda\right)$. The reader
will correctly infer from here that the following identity holds:
\begin{equation}
\sum_{\lambda\vdash m}d\left(\lambda\right)\mathcal{S}\left(\lambda,n\right)=n^{m}\,.
\end{equation}
For a fixed $\alpha$, $T_{\lambda\alpha j}$ transforms as the irreducible
representation $\lambda$ of $S_{m}$ when the $i_{x}$ indices are
permuted:
\begin{align}
T_{i_{1}i_{2}\cdots i_{m}} & \overset{\pi}{\rightarrow}T_{\pi\left(i_{1}i_{2}\cdots i_{m}\right)}\,,\\
T_{\lambda\alpha j} & \overset{\pi}{\rightarrow}\left[P_{\lambda}\left(\pi\right)\right]_{jj^{\prime}}T_{\lambda\alpha j^{\prime}}\,,
\end{align}
where $P_{\lambda}\left(\pi\right)$ stands for the matrices of the
irreducible representation $\lambda$ of $S_{m}$. This means that
the tensor $T$ contains $\mathcal{S}\left(\lambda,n\right)$ irreducible
representations $\lambda$ of $S_{m}$ (one for each value of $\alpha$).

All the above is true for any tensor. It is interesting now to consider
the particular scenario where there is not just one tensor, but rather
a list of tensors $T_{i_{1},\cdots,i_{m}}^{(1)}$, $T_{i_{1},\cdots,i_{m}}^{(2)}$,
..., $T_{i_{1},\cdots,i_{m}}^{\left(d(\mu)\right)}$ such that
\begin{equation}
T_{\pi\left(i_{1},\cdots,i_{m}\right)}^{(a)}=\left[P_{\mu}\left(\pi^{-1}\right)\right]_{ab}T_{i_{1},\cdots,i_{m}}^{(b)}\label{eq:30}
\end{equation}
for some partition $\mu$ of $m$; the $\phi^{4}$ interactions we
considered earlier are the motivation to study this kind of lists
of tensors (see text just below equation (\ref{eq:17}), where the
$T$'s were called $c$'s). Note that the $P$ matrices can always
be made unitary and real, so let us assume for simplicity that they
do have these properties. With the $\left(\lambda,\alpha,j\right)$
indices it is easy to see what is the most general form of the above
list of tensors; it must obey the constraint\footnote{Since $P_{\mu}\left(\pi^{-1}\right)$ is unitary and real, $\left[P_{\mu}\left(\pi^{-1}\right)\right]_{ab}=\left[P_{\mu}\left(\pi\right)\right]_{ba}$.}
\begin{equation}
\left[P_{\lambda}\left(\pi\right)\right]_{jj^{\prime}}T_{\lambda\alpha j^{\prime}}^{(a)}=\left[P_{\mu}\left(\pi\right)\right]_{ba}T_{\lambda\alpha j}^{(b)}
\end{equation}
and so, according to Shur's lemma from group theory, the components
of $T_{\lambda\alpha j}^{(a)}$ are non-zero only if $\lambda=\mu$
and $j=a$, plus their value is independent of $j$:
\begin{equation}
T_{\lambda\alpha j}^{(a)}=T\left(\alpha\right)\delta_{\lambda\mu}\delta_{aj}\,.
\end{equation}
The $T\left(\alpha\right)$ are free numbers; they are the free parameters
in the most general list of tensors $T^{(a)}$ subjected to the constraint
in equation (\ref{eq:30}), and there are $\mathcal{S}\left(\mu,n\right)$
of them since the index $\alpha$ can go from 1 to this number.

There is one final aspect of lists of tensors with permutation symmetries
which is worth looking at, as it will be relevant for the counting
of the minimum number of terms in an effective field theory. Consider
again a list of tensors which obeys equation (\ref{eq:30}) for some
partition $\mu$ of the integer $m$ ($\mu\vdash m$). This relation
states that if we have all the $T_{i_{1},\cdots,i_{m}}^{(a)}$ we
can reproduce the effect of swapping the indices $i_{x}$ by just
making linear combinations of the $T_{i_{1},\cdots,i_{m}}^{(a)}$
with the indices $i_{x}$ unchanged. What is noteworthy is that the
opposite is also true: the values of $T_{i_{1},\cdots,i_{m}}^{(a)}$
can be written down as a linear combination of $T_{\pi\left(i_{1},\cdots,i_{m}\right)}^{(1)}$
for different $\pi$ (the same is true if we used $T^{(2)}$ or any
other component of the list of tensors). This is a consequence of
the orthogonality relations among the entries of the unitary (and
real) matrices $P_{\mu}\left(\pi\right)$ of an irreducible representation:
\begin{equation}
T_{i_{1},\cdots,i_{m}}^{(a)}=\frac{d\left(\mu\right)}{m!}\sum_{\pi\in S_{m}}P_{\mu}\left(\pi\right)_{1a}T_{\pi\left(i_{1},\cdots,i_{m}\right)}^{(1)}\,.
\end{equation}
The precise form of this last expression is not particularly relevant;
rather, it suffices to keep in mind that all the entries of any of
the tensors $T^{(a)}$ can be obtained from the entries of a single
one of them ($T^{(1)}$ for example), given that there is the relation
(\ref{eq:30}).  However, we have started with a list of tensors
$T^{(a)}$ which, under permutations, transforms as an irreducible
representation $\mu$ of $S_{m}$. It is helpful to consider what
happens if instead of $\mu$ we had a reducible representation which
is a direct sum $\left(\mu_{1}\right)^{r_{1}}+\left(\mu_{2}\right)^{r_{2}}+\cdots$
of distinct irreducible parts $\mu_{i}$ repeated $r_{i}$ times.
It turns out that if each $\mu_{i}$ appears at most once in this
sum ($r_{i}=0$ or 1), then it is possible to pick a single combination
of the $T^{(a)}$ and generate all $T^{(a)}$ from it as we did before
with $T^{(1)}$. On the other hand, if there are $\mu_{i}$ which
appear more than once, $\max\left(r_{i}\right)$ linear combinations
of the $T^{(a)}$ are needed in order to generate all of them via
permutations of the $i_{1},\cdots,i_{m}$ indices.\footnote{These linear combinations of the $T^{(a)}$ must be carefully picked,
otherwise more than $\max\left(r_{i}\right)$ might be needed in order
to generate the full set of tensors $T^{(a)}$.}

~

In summary, two important facts stand out from this discussion about
tensors:
\begin{enumerate}
\item A list of tensors $T^{(a)}$ which transforms under some irreducible
representation $\mu$ when its indices are permuted [see equation
\ref{eq:30})] has $\mathcal{S}\left(\mu,n\right)$ independent entries,
where $n$ is the number of values each of the tensor indices can
take. For multiple irreducible representations $\mu_{i}$, the number
of independent entries is given simply by the sum of the numbers $\mathcal{S}\left(\mu_{i},n\right)$.
\item A list of tensors $T^{(a)}$ which transforms under a direct sum of
irreducible representations $\mu_{i}$, where each $\mu_{i}$ is repeated
$r_{i}$ times --- $\left(\mu_{1}\right)^{r_{1}}+\left(\mu_{2}\right)^{r_{2}}+\cdots$
--- can be reconstructed from just a few of the $T^{(a)}$. Specifically,
only $\max\left(r_{i}\right)$ linear combinations of the $T^{(a)}$
are needed for that.
\end{enumerate}
These two observations can be used to count the number of operators
and the minimum number of terms in an effective field theory.

\subsection{Application to operators with repeated fields}

We are now in a position to go back to the discussion of operators
with a repeated field $\chi$ which has two indices: a group index
$g$, and a flavor index $f$ going from 1 to some number $n$. The
product of $m$ $\chi$'s is either completely symmetric (if $\chi$
is a boson) or completely anti-symmetric (if $\chi$ is a fermion),

\begin{equation}
\chi_{g_{1}}^{f_{1}}\chi_{g_{2}}^{f_{2}}\cdots\chi_{g_{m}}^{f_{m}}\equiv\chi_{\left\{ g\right\} }^{\left\{ f\right\} }=\left(\pm1\right)^{\pi}\chi_{\pi\left\{ g\right\} }^{\pi\left\{ f\right\} }\,.
\end{equation}
In the Lagrangian, the $\left\{ g\right\} $ indices contract with
some numeric tensors $c_{\left\{ g\right\} }^{(\alpha)}$ determined
from group theory, while the $\left\{ f\right\} $ indices contract
with a parameter tensor $\kappa_{\left\{ f\right\} }^{(\alpha)}$
(such as the Yukawa matrices in the Standard Model). The extra $\alpha$
index is needed because there might be more than one $c_{\left\{ g\right\} }$
contraction allowed by the model's symmetries. We then have the expression
\begin{equation}
\mathscr{L}=\cdots+\kappa_{\left\{ f\right\} }^{(\alpha)}c_{\left\{ g\right\} }^{(\alpha)}\chi_{\left\{ g\right\} }^{\left\{ f\right\} }+\cdots\,.
\end{equation}
The numerical tensors $c_{\left\{ g\right\} }^{(\alpha)}$ always
obey a symmetry relation of the type
\begin{equation}
c_{\pi\left\{ g\right\} }^{(\alpha)}=\left[P\left(\pi^{-1}\right)\right]_{\alpha\beta}c_{\left\{ g\right\} }^{(\beta)}
\end{equation}
where the $P$ matrices form a representation (perhaps reducible)
of the permutation group $S_{m}$, and they can be made real and unitary.
As a consequence we see that the parameter tensor $\kappa_{\left\{ f\right\} }^{(\alpha)}$
possesses the following symmetry:
\begin{equation}
\kappa_{\pi\left\{ f\right\} }^{(\alpha)}=\left(\pm1\right)^{\pi}\left[P\left(\pi^{-1}\right)\right]_{\alpha\beta}\kappa_{\left\{ f\right\} }^{(\beta)}\,.
\end{equation}
The product $\left(\pm1\right)^{\pi}P$ is itself a representation
$\widehat{P}$ of $S_{m}$, so we can write
\begin{equation}
\kappa_{\pi\left\{ f\right\} }^{(\alpha)}=\left[\widehat{P}\left(\pi^{-1}\right)\right]_{\alpha\beta}\kappa_{\left\{ f\right\} }^{(\beta)}\,.
\end{equation}
This representation $\widehat{P}$ can be decomposed into irreducible
representations $\mu\vdash m$ of $S_{m}$: $\widehat{P}=\sum_{\mu\vdash m}r_{\mu}\mu$,
where $r_{\mu}$ is the multiplicity of $\mu$ in $\widehat{P}$.\footnote{As mentioned earlier, the notation $\mu\vdash m$ means that $\mu=\left\{ \mu_{1},\mu_{2},\cdots\right\} $
is a partition of the integer $m$ (so the sum of the $\mu_{i}$ adds
up to $m$). Therefore $\sum_{\mu\vdash m}$ represents a sum over
all partitions $\mu$ of $m$.} The main point of the earlier discussion on tensors is that $\kappa_{\left\{ f\right\} }^{(\alpha)}$
contains 
\begin{equation}
\sum_{\mu\vdash m}r_{\mu}\mathcal{S}\left(\mu,n\right)
\end{equation}
free parameters, so this is the total number of operators, and furthermore
they all can be written with
\begin{equation}
\max\left(r_{\mu}\right)
\end{equation}
terms.\footnote{Those $\mu$'s having a Young diagram with more rows than the available
flavors ($n$) should not be considered in this last expression; they
are too anti-symmetric and as a consequence they are not associated
with any operator [$\mathcal{S}\left(\mu,n\right)=0$].} These comments apply to operators $\chi^{m}$ with a single type
of field $\chi$; however the generalization is trivial for cases
where there are more fields, $\chi^{m}\chi^{\prime m^{\prime}}\chi^{\prime\prime m^{\prime\prime}}\cdots$ (namely
 the above considerations apply separately to each group of
repeated fields).

The fact that generally there is more than one group index $g$ is
not a problem either. For example, the Standard Model fields carry
an index from each of the four groups $SU(3)_{C}$, $SU(2)_{L}$,
$SU(2)_{l}$ and $SU(2)_{r}$. These indices contract with some symmetry
$P^{C}$, $P^{L}$, $P^{l}$ and $P^{r}$ so, taken together as a
single index $g$, the relevant matrix $P$ is given by the Kronecker
product $P^{C}\otimes P^{L}\otimes P^{l}\otimes P^{r}$, and $\widehat{P}=\left(\pm1\right)^{\pi}P^{C}\otimes P^{L}\otimes P^{l}\otimes P^{r}$.

~

Take the operators of type $QQQL$: there are $n_{L}n_{Q}\left(2n_{Q}^{2}+1\right)/3$
of them for $n_{Q,L}$ flavors of the fields $Q,L$, and they can
all be written with one term. These two conclusions can be obtained
by tracking the permutation symmetry of the contraction of the $Q$'s,
as shown in table \ref{tab:QQQL} ($L$ is not repeated, so it is
pointless to track its permutation symmetry, which is trivial). In
particular, the color indices of the $Q$'s contract anti-symmetrically
($\left\{ 1,1,1\right\} $), while the $SU(2)_{L}$ indices are contracted
with the mixed symmetry $\left\{ 2,1\right\}$ as we shall see in section \ref{sec:5}.\footnote{$L$ is an $SU(2)_L$ doublet, so $QQQ$ must transform as a doublet as well. Furthermore, the product of three $SU(2)_L$ doublets (one for each $Q$) decomposes as $\boldsymbol{2}+\boldsymbol{2}+\boldsymbol{4}$, and the pair of $\boldsymbol{2}$'s in this sum transforms as the 2-dimensional representation $\{2,1\}$ under $S_3$ permutations. This particular case is discussed at the beginning of section \ref{sec:5}.} The same is true
for the $SU(2)_{l}$ part of the Lorentz group. Since the $Q$'s are
left-handed fields, we could either ignore entirely the group $SU(2)_{r}$, or alternative
notice that three singlets of $SU(2)_{r}$ 
contract trivially in a symmetric fashion ($\left\{ 3\right\} $).
Finally, we should take into account that these are fermionic fields,
so they anti-commute ($\left\{ 1,1,1\right\} $). The full symmetry
under permutations of the $QQQ$ contractions is given by the product
of these $S_{3}$ representations ($\{3\}\times\left\{ 2,1\right\} ^{2}\times\left\{ 1,1,1\right\} ^{2}$),
which decomposes into the irreducible
components $\left\{ 3\right\} +\left\{ 2,1\right\} +\left\{ 1,1,1\right\} $.
Note that there are four different ways to contract $QQQL$ (due to
the existence of two different $SU(2)_{L}$ contractions as well as
two different contractions of the Lorentz indices), and indeed we
see that $d\left(\left\{ 3\right\} \right)+d\left(\left\{ 2,1\right\} \right)+d\left(\left\{ 1,1,1\right\} \right)=1+2+1=4$.
However, we do not need to write down 4 terms in the Lagrangian: since
each irreducible representation of $S_{3}$ appears only once, we infer
that a single term is enough. Furthermore, the total number of operators
of the type $QQQL$ is given by the number\ytableausetup{boxsize=0.6em}
\begin{equation}
\left[\mathcal{S}\left(\ydiagram{3},n_{Q}\right)+\mathcal{S}\left(\ydiagram{2,1},n_{Q}\right)+\mathcal{S}\left(\ydiagram{1,1,1},n_{Q}\right)\right]\mathcal{S}\left(\ydiagram{1},n_{L}\right)=\frac{n_{L}n_{Q}\left(2n_{Q}^{2}+1\right)}{3}\,.
\end{equation}
For $n_{Q}=n_{L}=3$, this adds up to 57 complex operators \cite{Alonso:2013hga}.
Historically, the dimension six baryon number violating operators
were written as 6 terms in \cite{Weinberg:1979sa} and \cite{Wilczek:1979hc},
although it is possible to do so with only 4 \cite{Abbott:1980zj}.
In the particular case of $QQQL$ type operators, they were written
as 2 terms in \cite{Weinberg:1979sa} and in \cite{Grzadkowski:2010es}\footnote{In the third arXiv version of this paper, only one term is used.}
but, as stated above, these operators require only 1 term. Notice
that Fierz identities were not used to reach this conclusion; spinors
and any other field with Lorentz indices are viewed as representations
of an $SU(2)_{l}\times SU(2)_{r}$ group, and in turn this group is
treated in exactly the same way as the gauge group.

\ytableausetup{boxsize=0.5em}
\begin{center}
\begin{table}[tbph]
\begin{centering}
\begin{tabular}{cccc}
\toprule 
 & $QQQ$ &  & $L$\tabularnewline
\midrule
$SU(3)_{C}$ & \ydiagram{1,1,1} &  & \ydiagram{1}\tabularnewline
$SU(2)_{L}$ & \ydiagram{2,1} &  & \ydiagram{1}\tabularnewline
$SU(2)_{l}$ & \ydiagram{2,1} &  & \ydiagram{1}\tabularnewline
$SU(2)_{r}$ & \ydiagram{3} &  & \ydiagram{1}\tabularnewline
Grassmann & \ydiagram{1,1,1} &  & \ydiagram{1}\tabularnewline
\midrule
Total symmetry & $\ydiagram{1,1,1}^{\;2}\times\ydiagram{2,1}^{\;2}\times\ydiagram{3}=\ydiagram{3}+\ydiagram{2,1}+\ydiagram{1,1,1}$ & \hspace{1cm} & $\ydiagram{1}^{\,5}=\ydiagram{1}$\tabularnewline
\bottomrule
\end{tabular}
\par\end{centering}
\caption{\label{tab:QQQL}Permutation symmetry of the quark fields $Q$ in
$QQQL$-type operators (since there is just one $L$, there is nothing
noteworthy about it). The 4 possible contractions of these fields
transform under $S_{3}$ permutations of the $Q$'s as the sum of
irreducible representations $\left\{ 3\right\} +\left\{ 2,1\right\} +\left\{ 1,1,1\right\} $
($\left\{ 2,1\right\} $ is a 2-dimensional representation of $S_{3}$,
so in total we do have 4 contractions). As explained in the text,
from this information we can readily conclude that there are 57 $QQQL$
operators, which can all be written as a single term in the Lagrangian.}
\end{table}
\par\end{center}

Consider another example: the interaction of 4 right-handed neutrinos
mentioned in the introduction ($N^{c}N^{c}N^{c}N^{c}$). Each neutrino
transforms as a doublet of the $SU(2)_{l}$ group (which is part of
the Lorentz group), and we have seen that the two invariant contractions
of four doublets have a permutation symmetry $\{2,2\}$. Furthermore,
the components of the $N^{c}$'s are anti-commuting fields, so there
is a total anti-symmetry $\left\{ 1,1,1,1\right\} $ to be taken into
consideration. Overall, under permutations, the $N^{c}N^{c}N^{c}N^{c}$
interactions have a symmetry
\begin{equation}
\ydiagram{2,2}\times\ydiagram{1,1,1,1}=\ydiagram{2,2}\,,
\end{equation}
meaning that for $n$ flavors there are
\begin{equation}
\mathcal{S}\left(\ydiagram{2,2},n\right)=\frac{n^{2}\left(n^{2}-1\right)}{12}
\end{equation}
complex operators of the form $N^{c}N^{c}N^{c}N^{c}$, which can all
be written down as a single Lagrangian term. This operator counting
matches the one obtained in \cite{Liao:2016qyd} by other means. More
generally, using the same arguments, one concludes that the interactions
of $2m$ right-handed neutrinos have a permutation symmetry
\begin{equation}
\underbrace{\left\{ 2,2,\cdots,2\right\} }_{m}
\end{equation}
and so there are
\begin{equation}
\mathcal{S}\left(\underbrace{\left\{ 2,2,\cdots,2\right\} }_{m},n\right)=\begin{cases}
\frac{n!\left(n+1\right)!}{m!\left(m+1\right)!\left(n-m+1\right)!\left(n-m\right)!} & n\geq m\\
0 & n<m
\end{cases}
\end{equation}
complex operators of the type $\left(N^{c}\right)^{2m}$, which again
can all be written down as a single term in the Lagrangian (for a
fixed value of $m$).

\section{\label{sec:4-derivatives}Operators with derivatives}

\subsection{\label{subsec:Handling-operator-redundancies}Handling operator redundancies}

The values of a function and the value of its derivatives do not need
to be correlated in any way, so a field $\psi$ and its space-time
derivatives $\partial^{n}\psi$ can be treated as different fields
altogether. There is however one problem: some operators obtained
in this way might be redundant (let us call them $\mathcal{O}_{i}^{\textrm{0}}$),
and two Lagrangians differing by such operators will be equivalent.

Faced with this situation, one should work with classes of equivalent
Lagrangians $\mathcal{C}_{\mathscr{L}}=\left\{ \mathscr{L}+\alpha_{i}\mathcal{O}_{i}^{0}\right\} $
rather than individual Lagrangians. But these $\mathcal{C}_{\mathscr{L}}$
are rather abstract constructions, so it might be better to look at
redundancies in effective fields theories as a linear algebra problem.
For a given model, the Lagrangian can be viewed as a vector in a vector
space spanned by some $z$ operators $\mathcal{O}_{1},\mathcal{O}_{2},\cdots,\mathcal{O}_{z}$,
yet there might exist $r$ combinations $\mathcal{O}_{i}^{0}=\sum_{j}\mathfrak{r}_{ij}\mathcal{O}_{j}$
which are null for practical effects. Hence, there are only $z-r$
meaningful operators $\mathcal{O}_{i}^{\slashed{0}}=\sum_{j}\mathfrak{m}_{ij}\mathcal{O}_{j}$
which can be selected in more than one way, as long as the block matrix
\begin{equation}
\left(\begin{array}{c}
\mathfrak{r}\\
\mathfrak{m}
\end{array}\right)
\end{equation}
has full rank.

Given $\mathfrak{r}$, the choice of matrix $\mathfrak{m}$ is not
unique, but there is a particularly simple possibility: we start with
$\mathfrak{m}_{ij}=\delta_{ij}$ and for each row $R$ of $\mathfrak{r}$
(let us call it $\mathfrak{r}_{R}$) a line $M$ of $\mathfrak{m}$
($\mathfrak{m}_{M}$) satisfying the condition $\mathfrak{r}_{R}\cdot\mathfrak{m}_{M}\neq0$
is removed. In practice, this corresponds to looking at the operators
$\mathcal{O}_{j}$ appearing in the redundancy $\sum_{j}\mathfrak{r}_{Rj}\mathcal{O}_{j}$
and dropping one of them from any further analysis (as if we were
solving the equation $\sum_{j}\mathfrak{r}_{Rj}\mathcal{O}_{j}=0$
for one of the $\mathcal{O}_{j}$'s).

This way of selecting non-redundant operators will always work when
there is a single redundancy; however, repeating this simple procedure
for multiple redundancies might be problematic, so these more complicated
scenarios require caution.\footnote{As an example, consider three operators $\mathcal{O}_{1,2,3}$. If
there is a redundancy $\mathcal{O}_{1}+\mathcal{O}_{2}+\mathcal{O}_{3}$
we might deal with it by dropping the operator $\mathcal{O}_{3}$,
in which case we would consider that the space of all non-equivalent
Lagrangians is generated by $\mathcal{O}_{1}$ and $\mathcal{O}_{2}$.
If there is a second redundancy $-\mathcal{O}_{1}+\mathcal{O}_{2}+\mathcal{O}_{3}$
one cannot drop $\mathcal{O}_{3}$ (since we have done so already),
but more importantly we cannot simply drop $\mathcal{O}_{2}$ either,
and say that $\mathcal{O}_{1}$ alone generates all non-equivalent
Lagrangians. That is because the vectors $\mathcal{O}_{1}+\mathcal{O}_{2}+\mathcal{O}_{3}$,
$-\mathcal{O}_{1}+\mathcal{O}_{2}+\mathcal{O}_{3}$ and $\mathcal{O}_{1}$
are not linearly independent.}  Reference \cite{Gripaios:2018zrz} is particularly relevant for
this linear algebra view on operator redundancies.

~

Two types of potentially redundant operators are often considered:
(1) those which are zero when the classical equations of motion (EOM)
of the fields are applied, and (2) those which are a divergence of
a vector field ($\partial_{\mu}\mathcal{O}^{\mu}$). Both of these
kinds of operators can be ignored only under some assumptions, whose
merits will not be assessed in this work. Rather, I will just mention
what needs to be done if one wants to factor out these operators.
The solution to these problems was already given in the papers \cite{Lehman:2015coa}
and \cite{Henning:2015alf}; the following two sub-sections elaborate
on the proposal in \cite{Lehman:2015coa} to remove EOM redundancies,
and they also contain a discussion of a simple adaptation to the procedure
mentioned in \cite{Henning:2015alf} for factoring out operators of
the type $\partial_{\mu}\mathcal{O}^{\mu}$.

It should be stressed at this point that derivatives in a gauge theory
always appear through the combination $\partial_{\mu}+igT^{a}A_{\mu}^{a}\equiv D_{\mu}$,
where $D_{\mu}$ is the well-known covariant derivative. For this
reason, all derivatives should be seen as being covariant. However,
because $\left[D_{\mu},D_{\nu}\right]$ can be written down with field
strength tensors $\mathcal{F}_{\mu\nu}$, in order not to over-count
operators one should only consider the completely symmetric part of
$D_{\mu_{1}}D_{\mu_{2}}\cdots$ applied to some field $\psi$, in
which case the $D_{\mu}$'s can be seen as commuting with each other
--- just as normal partial derivatives. To highlight that the term
$igT^{a}A_{\mu}^{a}$ is not important, in this work the symbol $\partial_{\mu}$
will be used instead of $D_{\mu}$.

\subsection{\label{subsec:Equations-of-motion}Equations of motion}

It can be shown that the classical equations of motion of a field
$\chi$ can be used to reduce the number of non-renormalizable operators
in an effective field theory \cite{Politzer:1980me,Arzt:1993gz}.
That is because non-renormalizable operators proportional to the quantity
\begin{equation}
\frac{\delta\mathscr{L}_{4}}{\delta\chi}-\partial_{\mu}\frac{\delta\mathscr{L}_{4}}{\delta\left(\partial_{\mu}\chi\right)}\label{eq:EOM}
\end{equation}
or its derivatives, where $\mathscr{L}_{4}$ is the renormalizable
part of a Lagrangian, do not affect the $S$-matrix. The standard
approach to these redundancies is to remove the operators with the
highest number of derivatives [the second term in expression (\ref{eq:EOM})],
since these can be traded by other operators with fewer derivatives
[the first term in expression (\ref{eq:EOM})].

The authors of \cite{Lehman:2015coa} pointed out that these redundancies
due to the equations of motion are best seen if we decompose the field
derivatives $\partial^{i}\chi$ into irreducible representations of
the Lorentz group.\footnote{The exponent $i$ in a derivative refers to the product of $\partial\times\partial\times\cdots$
with $i$ factors. Similarly, $\partial^{2}=\partial\times\partial$.
The exponent only refers to a space-time index if a Greek letter is
used ($\partial^{\mu}$ for instance).} For example, the 16 second derivatives of a scalar $\phi$ transform
under $SU(2)_{l}\times SU(2)_{r}$ as
\begin{equation}
\partial^{2}\phi=\left(0,0\right)+\left(0,1\right)+\left(1,0\right)+\left(1,1\right)\,,
\end{equation}
but apart from $\left(1,1\right)$ (with 9 components), these irreducible
representations of the Lorentz are redundant. The reason is the following.
The parts $\left(0,1\right)$ and $\left(1,0\right)$ are anti-symmetric
under permutation of the two derivatives, so they can be written with
the field stress tensors, $\left[\partial^{2}\phi\right]_{\left(0,1\right)+\left(1,0\right)}\propto\mathcal{F}\phi$,
so we should discard them. On the other hand the equation of motion
of $\phi$ equates $\left[\partial^{2}\phi\right]_{\left(0,0\right)}=\Box\phi$
to terms without derivatives so, if we follow the rule of keeping
terms with as few derivatives as possible, only the $\left(1,1\right)$
part of $\partial^{2}\phi$ needs to be kept.

All 4 first derivatives of $\phi$ which transform as 
\begin{equation}
\partial\phi=\left(\frac{1}{2},\frac{1}{2}\right)
\end{equation}
are to be kept as well, but in the case of a left-handed fermion $\psi$
the situation is different. The 8 components of $\partial\psi$ transform
as
\begin{equation}
\partial\psi=\left(1,\frac{1}{2}\right)+\left(0,\frac{1}{2}\right)\,,
\end{equation}
yet the equations of motion of $\psi$ relate $\left[\partial\psi\right]_{\left(0,\frac{1}{2}\right)}$
to a quantity with no derivatives, so we should keep only $\left[\partial\psi\right]_{\left(1,\frac{1}{2}\right)}$.
Adding one more derivative, in analogy to the scalar case, we should
worry only about $\left[\partial^{2}\psi\right]_{\left(\frac{3}{2},1\right)}$.

This procedure works well not just for Hilbert series calculations
\cite{Lehman:2015coa} but also for the more straightforward approach
to operator counting being described in this work. In practice, the
EOM degeneracies are taken into account by introducing a tower of
new fields representing the non-redundant parts of $\partial\chi,$$\partial^{2}\chi,\cdots,\partial^{i}\chi$,
for every standard model field $\chi$. They have the same gauge quantum
numbers as $\chi$ but different Lorentz representations. Since we
are expanding significantly the number of fields in the effective
field theory, this has the adverse effect of increasing the computational
complexity of the calculations. Nevertheless, the procedure to handle
the equations of motion is conceptually very simple.

~

In any case, one must know what are the components of the field derivatives
to be kept. For a generic field $\chi$ we just saw that there are
two considerations to have in mind:
\begin{itemize}
\item We want only the $\partial^{i}\chi$ components which cannot be written
with field strength tensors, hence they correspond to those components
which are completely symmetric under permutations of the derivatives.
We may represent them by the expression $\left\{ \partial^{i}\right\} \chi$
which, in general, transforms as a reducible representation of the
Lorentz group --- for example $\left\{ \partial^{4}\right\} \phi=\left(2,2\right)+\left(1,1\right)+\left(0,0\right)$.
\item Some components of $\left\{ \partial^{i}\right\} \chi$ are also
redundant for another reason. If we represent the equation of motion
of $\chi$ with the notation
\begin{equation}
\left\langle \partial^{x}\chi\right\rangle =\cdots\quad\textrm{(}x=1\textrm{ for fermions and }x=2\textrm{ for bosons),}
\end{equation}
the operator $\left\langle \partial^{x}\chi\right\rangle $ and its
derivatives should also be discarded.
\end{itemize}
However, there is a difficulty with this two-stage reduction of operators:
we cannot just take the irreducible Lorentz representations in $\left\{ \partial^{i}\right\} \chi$
and remove all those in $\partial^{i-x}\left\langle \partial^{x}\chi\right\rangle $,
because some of the latter components are anti-symmetric under permutations
of the derivatives, and therefore can be written with the field strength
tensor $\mathcal{F}_{\mu\nu}$. So in reality the terms to be retained
are
\begin{equation}
\left[\partial^{i}\chi\right]_{\textrm{keep}}=\left\{ \partial^{i}\right\} \chi\textrm{ minus the components in }\left[\partial^{i-x}\left\langle \partial^{x}\chi\right\rangle \right]_{\textrm{all }\partial\textrm{'s sym. contracted}}.
\end{equation}
Figuring out what are the components of $\partial^{i}\chi$ to keep
is not a trivial exercise; it turns out that for a scalar $\phi=\left(0,0\right)$,
a left-hand fermion $\psi=\left(\frac{1}{2},0\right)$ and a field
strength tensor $F=\left\{ 1,0\right\} $ (recall that $\mathcal{F}=F+F^{*}$)
only the highest spin component should be retained, i.e.
\begin{align}
\left[\partial^{i}\phi\right]_{\textrm{keep}} & =\left(\frac{i}{2},\frac{i}{2}\right)\,,\\
\left[\partial^{i}\psi\right]_{\textrm{keep}} & =\left(\frac{i+1}{2},\frac{i}{2}\right)\,,\\
\left[\partial^{i}F\right]_{\textrm{keep}} & =\left(\frac{i+2}{2},\frac{i}{2}\right)\,.
\end{align}
The rest of this subsection discusses this result (see also \cite{Henning:2017fpj}).

Let us consider first a scalar $\phi$. It is not hard to show that
\begin{align}
\left\{ \partial^{i}\right\} \phi & =\left(\frac{i}{2},\frac{i}{2}\right)+\left(\frac{i}{2}-1,\frac{i}{2}-1\right)+\left(\frac{i}{2}-2,\frac{i}{2}-2\right)+\cdots\,.\label{eq:Der^i-diphi}
\end{align}
On the other hand, in order to calculate the Lorentz transformation
properties of the components $\partial^{i-2}\left\langle \partial^{2}\phi\right\rangle $
which need to be removed, we may want to apply the $i-2$ derivatives
in a completely symmetric way:
\begin{align}
\left\{ \partial^{i-2}\right\} \left\langle \partial^{2}\phi\right\rangle  & =\left(\frac{i}{2}-1,\frac{i}{2}-1\right)+\left(\frac{i}{2}-2,\frac{i}{2}-2\right)+\cdots\,.
\end{align}
It is then tempting to subtract these Lorentz irreducible representations
from the ones in equation (\ref{eq:Der^i-diphi}) and conclude that
the only non-redundant piece of $\left\{ \partial^{i}\right\} \phi$
is the one transforming as $\left(\frac{i}{2},\frac{i}{2}\right)$.
The last statement is true, but one should keep in mind that this
argument only works because none of the components of $\left\{ \partial^{i-2}\right\} \left\langle \partial^{2}\phi\right\rangle $
is proportional to $\mathcal{F}_{\mu\nu}$; in other words, all components
of $\left\{ \partial^{i-2}\right\} \left\langle \partial^{2}\phi\right\rangle $
have a part which is completely symmetric under all permutations of
all $i$ derivatives.\footnote{We know that the two derivatives in $\left\langle \partial^{2}\phi\right\rangle =\eta^{\mu\nu}\partial_{\mu}\partial_{\nu}$
are contracted symmetrically, and so are the remaining ones in $\left\{ \partial^{i-2}\right\} $
(obviously). Therefore the $\left(i+1\right)!/\left[3!\left(i-2\right)!\right]$
components of $\left\{ \partial^{i-2}\right\} \left\langle \partial^{2}\phi\right\rangle $
transform as a representation $\left\{ i-2\right\} \times\left\{ 2\right\} $
of the group $S_{i-2}\times S_{2}$ under permutations of the derivatives.
From this information alone, one cannot conclude that under permutations
of the bigger group $S_{i}\supset S_{i-2}\times S_{2}$ all these
components transform as a fully symmetric representation $\left\{ i\right\} $
(it is conceivable that they could also transform as $\left\{ i-1,1\right\} $
or $\left\{ i-2,2\right\} $).}

For derivatives of a left-handed fermion, the transformation properties
under Lorentz transformations are as follows:
\begin{align}
\left\{ \partial^{i}\right\} \psi & =\left(\frac{i+1}{2},\frac{i}{2}\right)+\left(\frac{i+1}{2}-1,\frac{i}{2}-1\right)+\left(\frac{i+1}{2}-2,\frac{i}{2}-2\right)+\cdots\nonumber \\
 & +\left(\frac{i-1}{2},\frac{i}{2}\right)+\left(\frac{i-1}{2}-1,\frac{i}{2}-1\right)+\left(\frac{i-1}{2}-2,\frac{i}{2}-2\right)+\cdots\,,\label{eq:der^i psi}\\
\left\{ \partial^{i-1}\right\} \left\langle \partial\psi\right\rangle  & =\left(\frac{i+1}{2}-1,\frac{i}{2}-1\right)+\left(\frac{i+1}{2}-2,\frac{i}{2}-2\right)+\cdots\nonumber \\
 & +\left(\frac{i-1}{2},\frac{i}{2}\right)+\left(\frac{i-1}{2}-1,\frac{i}{2}-1\right)+\left(\frac{i-1}{2}-2,\frac{i}{2}-2\right)+\cdots\,.\label{eq: der-EOM-psi}
\end{align}
Assuming again that all components of $\left\{ \partial^{i-1}\right\} \left\langle \partial\psi\right\rangle $
are, at least in part, fully symmetric, the only non-redundant component
of $\partial^{i}\psi$ is $\left(\frac{i+1}{2},\frac{i}{2}\right)$,
with maximum spin. This is the only part of the right-hand side of
equation (\ref{eq:der^i psi}) which remains after subtracting the
right-hand side of equation (\ref{eq: der-EOM-psi}).

\begin{figure}
\begin{centering}
\includegraphics[scale=1.8]{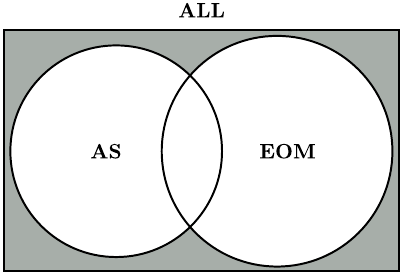}
\par\end{centering}
\caption{\label{fig:Venn-diagram}The set of all $\left\{ \partial^{i}\right\} \mathcal{F}$
components (\textit{ALL}) includes components which are redundant
due to anti-symmetric contractions of the $i+1$ derivatives (set
\textit{AS}) and/or due to the equations of motion (set \textit{EOM}).
In order to remove these redundancies and obtain the truly important
$\left\{ \partial^{i}\right\} \mathcal{F}$ components (gray part
of the diagram), one must understand which elements of \textit{ALL}
are in both\textit{ AS} and \textit{EOM}. (For $i=6$, the diagram
shown here faithfully represents the relative size of the sets and
their intersections.)}
\end{figure}

Finally, we have to consider what happens to derivatives of field
strength tensors $\mathcal{F}$. As before, one might try to compute
the irreducible Lorentz representations associated with the components
$\left\{ \partial^{i}\right\} \mathcal{F}$ (let us call these components
\textit{ALL}) and remove those which appear in the equations of motion
$\left\{ \partial^{i-1}\right\} \left\langle \partial\mathcal{F}\right\rangle $
(we may call this set of components \textit{EOM}). The trouble is
that in some cases the $i+1$ derivatives in $\left\{ \partial^{i}\right\} \mathcal{F}$
and $\left\{ \partial^{i-1}\right\} \left\langle \partial\mathcal{F}\right\rangle $\footnote{Note that $\mathcal{F}$ already contains one derivative.}
are not contracted in a fully symmetric way (they form a set \textit{AS}).
Given that all derivatives are in reality covariant, these $\left\{ \partial^{i}\right\} \mathcal{F}$
components can be written with two or more field strength tensors
$\mathcal{F}$, so they are redundant. We wish to calculate the elements
of \textit{ALL} which are neither in \textit{EOM} nor in \textit{AS},
but this is complicated by the fact that these last two sets intersect
(see figure \ref{fig:Venn-diagram}). The number of elements of each
set, as a function of $i$, is the following:
\begin{align}
6\mathcal{S}(\left\{ i\right\} ,4) & =\left(i+3\right)\left(i+2\right)\left(i+1\right)\quad\left(\#ALL\right),\label{eq:numb-all-derF}\\
\mathcal{S}(\left\{ i,1,1\right\} ,4) & =\frac{\left(i+3\right)\left(i+1\right)i}{2}\quad\quad\quad\;\left(\#AS\right)\,,\\
4\mathcal{S}(\left\{ i-1\right\} ,4) & =\frac{2\left(i+2\right)\left(i+1\right)i}{3}\quad\quad\;\;\left(\#EOM\right)\,.
\end{align}
The sets \textit{EOM} and \textit{AS} cannot be disjoint simply from
the fact that $\#AS+\#EOM$ can be larger than $\#ALL$. Up to large
values of $i$, it is possible to check explicitly that the intersection
of \textit{EOM} and \textit{AS} contains
\begin{equation}
\frac{\left(i+1\right)i\left(i-1\right)}{6}\quad\left[\#\left(AS\cap EOM\right)\right]
\end{equation}
 elements, and from here we can conclude that there are
\begin{equation}
2\left(i+3\right)\left(i+1\right)\quad\left\{ \#\left[ALL\setminus\left(AS\cup EOM\right)\right]\right\} \label{eq:numb-keep-derF}
\end{equation}
components of $\left\{ \partial^{i}\right\} \mathcal{F}$ which are
not redundant. The maximum spin components, transforming under the
Lorentz representation $\left(\frac{i+2}{2},\frac{i}{2}\right)+\left(\frac{i}{2},\frac{i+2}{2}\right)$,
are certainly not contained in either \textit{EOM} nor \textit{AS},
and they are precisely $2\left(i+3\right)\left(i+1\right)$ in total.
Hence, one can conclude that these are the only components of $\left\{ \partial^{i}\right\} \mathcal{F}$
to keep.

\subsection{\label{subsec:Integration-by-parts}Integration by parts}

There is another kind of operator redundancy. The action is calculated
by integrating the Lagrangian density over all spacetime, so what
matters are not the operators $\mathcal{O}$ themselves but rather
their spacetime integral. If $\mathcal{O}$ is a divergence, $\mathcal{O}=\partial^{\mu}\mathcal{O}_{\mu}$,
and the value of $\mathcal{O}_{\mu}$ drops to zero fast enough at
large distances/times, then by the divergence theorem
\begin{equation}
\int_{\mathcal{M}}\mathcal{O}\,d^{4}x=0\,,
\end{equation}
where $\mathcal{M}$ is some infinitely large volume of spacetime.
Putting aside the possibility that the above integral might not always
vanish, operators of the form $\mathcal{O}=\partial^{\mu}\mathcal{O}_{\mu}$
should be discarded. This is often seen as an \textit{integration
by parts} redundancy: if the sum of two operators $\mathcal{O}_{\mu}^{(1)}$
and $\mathcal{O}_{\mu}^{(2)}$ is a total divergence, an integration
by parts reveals that $\mathcal{O}_{\mu}^{(1)}$ is equivalent to
$-\mathcal{O}_{\mu}^{(2)}$.

We may track the operators which are total derivatives by introducing
a dummy field $\mathcal{D}$ in the effective field theory representing
total derivatives, so that operators of the form $\mathcal{D}\left(\cdots\right)$
should be factored out. However, some of these latter operators are
redundant \cite{Henning:2015alf}, because there are linear combinations
of them which add up to zero. Unfortunately, it does not seem easy
to detect these relations among operators which are total divergences.

Nevertheless, the authors of \cite{Henning:2015alf} managed to count
them with the Hilbert series method and conformal field theory; they
also offered an interesting interpretation of their calculation which
does not rely on these particular theoretical tools. It goes as follows,
using the language of differential forms. The redundant operators
$\mathcal{O}=\partial^{\mu}\mathcal{O}_{\mu}$ are associated with those
differential 4-forms $\omega^{(4),\textrm{red}}$ which are exact,
meaning that they are the differential of some 3-form $\omega^{(3)}$:
$\omega^{(4),\textrm{red}}=d\omega^{(3)}$. Nevertheless, one cannot
simply consider all $\omega^{(3)}$ because the differential of some
3-forms is identically zero, i.e. $d\omega^{(3),\textrm{red}}=0$.
In turn, according to Poincaré's lemma, these `redundancies of the
redundancies' are associated with all those 3-forms which are the differentials
of a 2-form $\omega^{(2)}$: $\omega^{(3),\textrm{red}}=d\omega^{(2)}$.
This recursive argument goes on, but it eventually stops because of
the dimensionality of spacetime (four) when we reach 0-forms, which
are pure functions.

Let us rephrase this argument in the language of operators. In general,
a differential $i$-form $\omega^{(i)}$ can be written as
\begin{equation}
\omega^{(i)}=\mathcal{O}_{\mu_{1}\mu_{2}\cdots\mu_{i}}dx^{\mu_{1}}\wedge dx^{\mu_{2}}\dots\wedge dx^{\mu_{i}}\,,
\end{equation}
where $\mathcal{O}_{\mu_{1}\mu_{2}\cdots\mu_{i}}$ is a completely
anti-symmetric tensor under an exchange of indices; to highlight this
fact, from now on square brackets will be applied to its indices.
Applying the differential operator $4-i$ times to $\omega^{(i)}$
yields
\begin{equation}
d^{4-i}\omega^{(i)}\propto\partial^{\mu_{1}}\partial^{\mu_{2}}\cdots\partial^{\mu_{i}}\mathcal{O}_{\left[\mu_{1}\mu_{2}\cdots\mu_{i}\right]}dx^{0}\wedge dx^{1}\wedge dx^{2}\wedge dx^{3}\,.
\end{equation}
If $i>1$, the expressions $\partial^{\mu_{1}}\partial^{\mu_{2}}\cdots\partial^{\mu_{i}}\mathcal{O}_{\left[\mu_{1}\mu_{2}\cdots\mu_{i}\right]}$
are identically zero, but even so it is crucial that we keep track
of them. Consider the set $\left\{ \mathcal{O}\right\} $ of all scalar
operators allowed by the model's symmetries, as well as the set of
all operators $\left\{ \mathcal{O}_{\left[\mu_{1}\mu_{2}\cdots\mu_{i}\right]}\right\} $
with $i=1,2,3,4$ free spacetime indices, which are completely anti-symmetric
under the exchange of these indices. The non-redundant operators are
those in the set
\begin{equation}
\left\{ \mathcal{O}\right\} \setminus\left(\left\{ \partial^{\mu}\mathcal{O}_{\mu}\right\} \setminus\left(\left\{ \partial^{\mu}\partial^{\nu}\mathcal{O}_{\left[\mu\nu\right]}\right\} \setminus\left(\left\{ \partial^{\mu}\partial^{\nu}\partial^{\rho}\mathcal{O}_{\left[\mu\nu\rho\right]}\right\} \setminus\left\{ \partial^{\mu}\partial^{\nu}\partial^{\rho}\partial^{\sigma}\mathcal{O}_{\left[\mu\nu\rho\sigma\right]}\right\} \right)\right)\right)\,.\label{eq:IBP-non-redundant-operators}
\end{equation}

In summary, all linear combinations of the operators $\left\{ \mathcal{O}\right\} $
are non-zero, but we wish to remove the degeneracies $\left\{ \partial^{\mu}\mathcal{O}_{\mu}\right\} $.
Unfortunately, this last set includes some operators which are identically
zero, hence the need to include the remaining sets in expression (\ref{eq:IBP-non-redundant-operators}),
containing only null operators. If all we want is to count the number
of non-redundant operators up to some dimension $d$, then the result
is\footnote{In 4-dimensional spacetime, there is a one-to-one correspondence between
the

operators $\mathcal{O}_{\mu}$ and $\mathcal{O}_{\left[\mu\nu\rho\right]}$
with the same dimension, as well as between the operators $\mathcal{O}$
and $\mathcal{O}_{\left[\mu\nu\rho\sigma\right]}$. Therefore expression
(\ref{eq:IBP-number-operators}) is the same as $\left(\#\mathcal{O}^{dim\leq d}\right)-\left(\#\mathcal{O}_{\mu}^{dim\leq d-1}\right)+\left(\#\mathcal{O}_{\left[\mu\nu\right]}^{dim\leq d-2}\right)-\left(\#\mathcal{O}_{\mu}^{dim\leq d-3}\right)+\left(\#\mathcal{O}^{dim\leq d-4}\right)$.}
\begin{equation}
\left(\#\mathcal{O}^{dim\leq d}\right)-\left(\#\mathcal{O}_{\mu}^{dim\leq d-1}\right)+\left(\#\mathcal{O}_{\left[\mu\nu\right]}^{dim\leq d-2}\right)-\left(\#\mathcal{O}_{\left[\mu\nu\rho\right]}^{dim\leq d-3}\right)+\left(\#\mathcal{O}_{\left[\mu\nu\rho\sigma\right]}^{dim\leq d-4}\right)\,.\label{eq:IBP-number-operators}
\end{equation}

Each of the five numbers in this sum can be calculated by directly counting how many operators there are with the indicated transformation
property under the Lorentz group: Under $SU(2)_l\times SU(2)_r$, an operator with no subscript must transform as $\left(0,0\right)$, while those with the subscripts $\mu$, $\left[\mu\nu\right]$, $\left[\mu\nu\rho\right]$ and $\left[\mu\nu\rho\sigma\right]$ transform as
$\left(\frac{1}{2},\frac{1}{2}\right)$, $\left(1,0\right)+\left(0,1\right)$,
$\left(\frac{1}{2},\frac{1}{2}\right)$ and $\left(0,0\right)$, respectively.

Alternatively, one can avoid handling operators which are not
scalars by introducing the derivative $\mathcal{D}$ as a stand-alone
field as mentioned earlier, with all the expected properties, except
that it anti-commutes (making it a Grassmann field). With this dummy
field,

\begin{align}
\#\mathcal{O}_{\mu}^{dim\leq d-1} & =\#\left(\mathcal{D}^{\mu}\mathcal{O}_{\mu}\right)^{dim\leq d}\\
\#\mathcal{O}_{\left[\mu\nu\right]}^{dim\leq d-2} & =\#\left(\mathcal{D}^{\mu}\mathcal{D}^{\nu}\mathcal{O}_{\left[\mu\nu\right]}\right)^{dim\leq d}\\
\#\mathcal{O}_{\left[\mu\nu\rho\right]}^{dim\leq d-3} & =\#\left(\mathcal{D}^{\mu}\mathcal{D}^{\nu}\mathcal{D}^{\rho}\mathcal{O}_{\left[\mu\nu\rho\right]}\right)^{dim\leq d}\\
\#\mathcal{O}_{\left[\mu\nu\rho\sigma\right]}^{dim\leq d-4} & =\#\left(\mathcal{D}^{\mu}\mathcal{D}^{\nu}\mathcal{D}^{\rho}\mathcal{D}^{\sigma}\mathcal{O}_{\left[\mu\nu\rho\sigma\right]}\right)^{dim\leq d}
\end{align}
where the $\mathcal{O}$'s are assumed not to contain any $\mathcal{D}$. Due to these identities, it becomes necessary to track only the scalar operators up to dimension
$d$, and expression \eqref{eq:IBP-number-operators} now reads
\begin{equation}
\sum_{i=0}^{4}\left(-1\right)^{i}\left(\#\mathcal{O}^{dim\leq d}\textrm{ with }i\textrm{ }\mathcal{D}'s\right)\,.
\end{equation}
Note that this only works if $\mathcal{D}$ is a Grassmann field, as the subscripts ${\left[\mu\nu\cdots\right]}$ of the operators in expression \eqref{eq:IBP-number-operators} are completely anti-symmetrized. In fact, if $\mathcal{D}$ was a commuting field, then $\mathcal{D}^{\mu}\mathcal{D}^{\nu}\cdots\mathcal{O}_{\left[\mu\nu\cdots\right]}$ would be identically 0.

 As an example, table
\ref{tab:example-singlet} lists all types of operators with 4 derivatives
and 4 gauge invariant scalars $S$. From the permutation symmetries
of $S$, $\partial S$ and $\partial^{2}S$, we can compute that there
are the following numbers of operators, assuming $n$ flavors of $S$
(see section \ref{sec:3}):
\begin{alignat}{2}
\#\mathcal{O}= & \frac{1}{12}n^{2}\left(11n^{2}+12n+13\right) & \quad\quad & \left(\textrm{operators with }0\textrm{ }\mathcal{D}\textrm{'s}\right)\,,\\
\#\mathcal{O}_{\mu}= & \frac{1}{6}n^{2}\left(7n^{2}+3n+2\right) &  & \left(\textrm{operators with }1\textrm{ }\mathcal{D}\right)\,,\\
\#\mathcal{O}_{\left[\mu\nu\right]}= & \frac{1}{2}n^{2}\left(n^{2}-1\right) &  & \left(\textrm{operators with }2\textrm{ }\mathcal{D}\textrm{'s}\right)\,,\\
\#\mathcal{O}_{\left[\mu\nu\rho\right]}= & \frac{1}{6}n^{2}(n+1)(n+2) &  & \left(\textrm{operators with }3\textrm{ }\mathcal{D}\textrm{'s}\right)\,,\\
\#\mathcal{O}_{\left[\mu\nu\rho\sigma\right]}= & \frac{1}{24}n(n+1)(n+2)(n+3) &  & \left(\textrm{operators with }4\textrm{ }\mathcal{D}\textrm{'s}\right)\,.
\end{alignat}
It follows that there is a total of $\frac{1}{8}n\left(n^{3}+2n^{2}+3n+2\right)$
operators of the generic form $\partial^{4}S^{4}$.

\ytableausetup{boxsize=0.5em}

\begin{table}[tbph]
\begin{centering}
\begin{tabular}{ccc}
\toprule 
\#$\mathcal{D}$'s & Operator type & Symmetry of the fields $\left(S,\partial S,\partial^{2}S\right)$\tabularnewline
\midrule
\multirow{3}{*}{0} & $SS\left(\partial^{2}S\right)\left(\partial^{2}S\right)$ & $\left(\ydiagram{2},-,\ydiagram{2}\right)$\tabularnewline
 & $S\left(\partial S\right)\left(\partial S\right)\left(\partial^{2}S\right)$ & $\left(\ydiagram{1},\ydiagram{2},\ydiagram{1}\right)$\tabularnewline
 & $\left(\partial S\right)\left(\partial S\right)\left(\partial S\right)\left(\partial S\right)$ & $\left(-,\ydiagram{4},-\right)+\left(-,\ydiagram{2,2},-\right)+\left(-,\ydiagram{1,1,1,1},-\right)$\tabularnewline
\midrule
\multirow{2}{*}{1} & $\mathcal{D}SS\left(\partial S\right)\left(\partial^{2}S\right)$ & $\left(\ydiagram{2},\ydiagram{1},\ydiagram{1}\right)$\tabularnewline
 & $\mathcal{D}S\left(\partial S\right)\left(\partial S\right)\left(\partial S\right)$ & $\left(\ydiagram{1},\ydiagram{3},-\right)$+$\left(\ydiagram{1},\ydiagram{2,1},-\right)$+$\left(\ydiagram{1},\ydiagram{1,1,1},-\right)$\tabularnewline
\midrule
2 & $\mathcal{D}\mathcal{D}SS\left(\partial S\right)\left(\partial S\right)$ & $2\left(\ydiagram{2},\ydiagram{1,1},-\right)$\tabularnewline
\midrule
3 & $\mathcal{D}\mathcal{D}\mathcal{D}SSS\left(\partial S\right)$ & $\left(\ydiagram{3},\ydiagram{1},-\right)$\tabularnewline
\midrule
4 & $\mathcal{D}\mathcal{D}\mathcal{D}\mathcal{D}SSSS$ & $\left(\ydiagram{4},-,-\right)$\tabularnewline
\bottomrule
\end{tabular}
\par\end{centering}
\caption{\label{tab:example-singlet}Relevant types of operators of the generic
form $\partial^{4}S^{4}$, where $S$ is a scalar with trivial gauge
quantum numbers. As explained in the main text, their total number
is obtained by counting the operators with no $\mathcal{D}$'s and
subtracting those with one $\mathcal{D}$. However, this naive operation
subtracts too many operators. To compensate for this, one must put
back the operators with two $\mathcal{D}$'s, remove those with three
$\mathcal{D}$'s and finally put back those with four $\mathcal{D}$'s.}
\end{table}

Crucially, in order to arrive at this counting, one must be able to compute
the permutation symmetries shown in table \ref{tab:example-singlet}. This can
be done as discussed previously, and to demonstrate it consider, for
example, the operators of the type $\left(\partial S\right)^{4}$.
The field $\partial S$ is gauge invariant, but it transforms under the Lorentz group
$SU(2)_{l}\times SU(2)_{r}$ as a bi-doublet. We have already seen
that the $SU(2)$ invariant contractions of four doublets transform
under $S_{4}$ permutations as $\left\{ 2,2\right\} $ (see also the next section), hence this
statement holds for both $SU(2)_{l}$ and $SU(2)_{r}$. Taking into account these two groups,
 we conclude that the contractions
of the four $\partial S$'s transform under the representation $\left\{ 2,2\right\} \times\left\{ 2,2\right\} $,
and this product decomposes into the sum $\left\{ 4\right\} +\left\{ 2,2\right\} +\left\{ 1,1,1,1\right\} $
of $S_4$ irreducible representations, in agreement with table \ref{tab:example-singlet}.

As a second example, we shall look into those operators of type $\mathcal{D}\mathcal{D}SS\left(\partial S\right)\left(\partial S\right)$.
The two $\mathcal{D}$'s must contract symmetrically because there
is just one copy of this dummy field. Each $\mathcal{D}$ is a bi-doublet
of $SU(2)_{l}\times SU(2)_{r}$, so we are looking at the product
$\left(\frac{1}{2},\frac{1}{2}\right)\times\left(\frac{1}{2},\frac{1}{2}\right)=\left(1,1\right)_{S}+\left(0,0\right)_{S}+\left(1,0\right)_{A}+\left(0,1\right)_{A}$.
Given that $\mathcal{D}$ is a Grassmann field, we infer that $\mathcal{D}\mathcal{D}$
transforms as $\left(1,0\right)+\left(0,1\right)$. Therefore $\left(\partial S\right)\left(\partial S\right)$
must transform as either $\left(1,0\right)$ or $\left(0,1\right)$
--- these are two valid possibilities, and crucially in both case
the $\partial S$'s contract anti-symmetrically. The $S$'s
always contract in a symmetric way (after all, $S$ is a Lorentz and gauge invariant field), therefore we deduce that
for the two possible contractions of the Lorentz indices, the relevant permutation symmetry of  $\left(S,\partial S\right)$ is 
 $\left(\left\{ 2\right\} ,\left\{ 1,1\right\} \right)$.

\section{\label{sec:5}Discussion of the algorithm and some results}

\subsection{Implementation in a computer code}

The approach described in this work makes it possible to count operators
of an effective field theory, and also to extract some extra information
about them. However, in order to implement it, for a representation
$R$ of some Lie group $G$ one must be able to decompose the tensor
product $R^{m}=R\times R\times\cdots\times R$ into irreducible representations
of $G\times S_{m}$. It is not enough to know the decomposition of
such products into irreducible representations of the Lie group $G$
alone.

There is a widely known technique involving Young tableaux to extract
this information when $G$ is a special unitary group $SU(p)$ and
$R$ is its fundamental representation. Any $SU(p)$ representation
can be labeled by a partition or a Young diagram, much like the representations
of the permutation group, with two caveats:
\begin{enumerate}
\item Columns with $p$ rows can be ignored, so two Young diagrams differing
only by such columns stand for the same $SU(p)$ representation;
\item Diagrams with more than $p$ rows are not associated with any $SU(p)$
representation.
\end{enumerate}
If $R$ is the fundamental representation of $SU(p)$, it turns out
that the tensor product $R^{m}=R\times R\times\cdots\times R$ decomposes
into the sum of all representations $\lambda_{\lambda}$ of the group
$SU(p)_{S_{m}}$ ($\lambda$ is a partition of $m$ with at most $p$
rows). For example, in the case of four $SU(2)$ doublets, $\lambda$
can be $\left\{ 4\right\} $, $\left\{ 3,1\right\} $ or $\left\{ 2,2\right\} $,
so we obtain the decomposition
\begin{equation}
\boldsymbol{2}\times\boldsymbol{2}\times\boldsymbol{2}\times\boldsymbol{2}=\boldsymbol{5}_{\left\{ 4\right\} }+\boldsymbol{3}_{\left\{ 3,1\right\} }+\boldsymbol{1}_{\left\{ 2,2\right\} }\,,\label{eq:four-doublets}
\end{equation}
because the partitions $\lambda=\left\{ \lambda_{1}\right\} $ and
$\lambda=\left\{ \lambda_{1},\lambda_{2}\right\} $ correspond to
the $\left(\lambda_{1}+1\right)$- and $\left(\lambda_{1}-\lambda_{2}+1\right)$-dimensional
representations of $SU(2)$, respectively. Note that $\left\{ 4\right\} $,
$\left\{ 3,1\right\} $ or $\left\{ 2,2\right\} $ are irreducible
representations of $S_{4}$ with dimensions 1, 3 and 2, respectively.
So on the right-hand side of equation (\ref{eq:four-doublets}) there
is a total of $1\times5+3\times3+2\times1=16=2^{4}$ components, as
expected. One can alternatively express the product of four doublets
as
\begin{equation}
\boldsymbol{2}\times\boldsymbol{2}\times\boldsymbol{2}\times\boldsymbol{2}=\boldsymbol{5}+\boldsymbol{3}+\boldsymbol{3}+\boldsymbol{3}+\boldsymbol{1}+\boldsymbol{1}\,,
\end{equation}
but in doing so we are erasing critical information from equation
(\ref{eq:four-doublets}).

When discussing previously the $QQQL$ operator, it was stated that
the $SU(2)_L$ indices of the $Q$'s transform as $\{2,1\}$ under $S_3$ permutations.
I will now mention two proofs of this statement. On the one hand we just saw that the two
singlets in the product of fours doublets of $SU(2)$ transform as $\{2,2\}$ under $S_4$.
However, in $QQQL$ we are dealing with 3 equal doublets ($Q$) plus a fourth doublet ($L$) which is
a distinct field. We therefore do not care about all $4!=24$ permutations of the four doublets
but rather we are only interested in the 6 permutation of the subgroup $S_{3}\times S_{1}\subset S_{4}$.
It turns out that the irreducible representation $\{2,2\}$ of $S_4$
transforms as $\{2,1\}\times\{1\}$ when restricted to this subgroup, or simply $\{2,1\}$ if we ignore the trivial $S_1$ group.

An alternative method of arriving at this conclusion is to repeat the decomposition show in equation \eqref{eq:four-doublets},
but this time for three doublets only. With the same arguments as before, we get that
\begin{equation}
\boldsymbol{2}\times\boldsymbol{2}\times\boldsymbol{2}=\boldsymbol{4}_{\{3\}}+\boldsymbol{2}_{\{2,1\}}\,.
\end{equation}
Each of the three doublets here stands for one of the $Q$'s, and in order for $QQQ$ to form an $SU(2)_L$
invariant with $L$, which is also a doublet, we must select only the doublet contraction of $QQQ$, which
transforms as  $\{2,1\}$ under $S_3$ permutations.

The tight connection between $SU(p)$ and the permutation group which
was just mentioned is all that is needed to count operators of some
models (including SMEFT). However, it is worth pointing out that this
type of decomposition can be calculated for any representation of
any group, including discrete ones. The \texttt{LiE} program \cite{LieProgram}
does this efficiently for any simple Lie group, referring to it as
a computation of a \textit{plethysm}. The algorithm described in \texttt{LiE}'s
manual was implemented in the \texttt{Susyno} package for Mathematica
\cite{Fonseca:2011sy}. This latter program also contains several
other functions related to Lie algebras and the permutation group
$S_{n}$.\footnote{The group theory part of this code has recently been separated and enlarged in a separate package called \texttt{GroupMath} \cite{Fonseca:GroupMath}. All the computations involving Lie groups and/or the permutation group $S_n$ which are mentioned in this work can be performed with it.} The first version of the \texttt{Sym2Int} package \cite{Fonseca:2017lem}
uses them to list the operators without derivatives (nor field strength
tensors) of any effective field theory. This can be done up to some
arbitrary mass dimension. Adapting the solutions in the literature
\cite{Lehman:2015coa,Henning:2015alf} to tackle the problems inherent
to derivatives, the latest version of \texttt{Sym2Int} can list automatically
all operators in an effective field theory, including those with derivatives
and field strength tensors.

\subsection{Comparison with other methods and computer codes}

All SMEFT operators up to dimension 15 were counter and characterized
in the way described here in under two hours on a standard laptop
computer (details of these results can be found below, at the end
of this section). All the numbers and expressions given in \cite{Henning:2015alf}
(and in the accompanying data files) were successfully reproduced
with the alternative method described here.

Two other noteworthy codes should be mentioned. The first one is \texttt{DEFT}
\cite{Gripaios:2018zrz}: rather than just calculate the number of
operators and their symmetries, this code works with the actual operators,
performing the gauge and Lorentz index contractions explicitly. Dealing
with operators redundancies associated with derivatives then becomes
a linear algebra problem. Having the operators written down explicitly
is clearly something very useful for model builders and it opens up
several possibilities, such as of implementing in different ways the
operator redundancies discussed previously. One downside of the more
ambitious scope of \texttt{DEFT} is that it takes more time to perform
these calculations --- as a point of reference, the authors of \cite{Gripaios:2018zrz}
were able to calculate SMEFT up to dimension 8 for one fermion generation,
which corresponds to roughly a third of the number of SMEFT operators
up to dimension 6 ,with three generations.

The second noteworthy code is \texttt{BasisGen} \cite{Criado:2019ugp}.
It counts operators, and does so with the basic approach of multiplying
all fields together, and retaining only the gauge and Lorentz invariant
solutions. This is essentially the same method of counting operators
as the one described in this paper and implemented in \texttt{Sym2Int}
(this was called the traditional method in section \ref{sec:2}).
However, repeated fields are handled differently, and this has significant
impact on the computational time for models with multiple flavors.
Instead of $n$ flavors of a field $\psi$, \texttt{BasisGen} considers
$n$ distinct fields fields $\psi$, $\psi^{\prime}$, $\psi^{\prime\prime}$,
... with the same quantum numbers, meaning that flavor indices are
expanded. From this perspective all fields have a single flavor so,
referring back to the discussion in section \ref{sec:3}, one only
needs to retain the completely symmetric contractions of repeated
fields. Because each entry of the flavor tensor in front of a term
is evaluated separately, the computational time of this approach increases
sharply with the number of flavors. Concerning derivatives, \texttt{BasisGen}
deals with the integration-by-parts redundancies in a way which is
different than the one described in this work; however, it is likely
that the two methods are equivalent.

~

\subsection{Information beyond the number of operators}

It is possible to track each field flavor separately, both with the
traditional method described here (see also \cite{Criado:2019ugp}),
and with the Hilbert series method (this leads to a so-called multigraded
Hilbert series). However, doing so is very time consuming; therefore
in the following I will analyze the information which can be extracted
from both methods when the field flavor indices are not expanded.

The Hilbert series approach counts operators of a certain type for
an arbitrary number of generations $n$, telling us, for example,
that there are $n\left(n+1\right)/2$ operators of the type $LLHH$.
At first sight, the traditional method might seem to yield more information:
it computes that the $n\times n$ flavor matrix $\kappa_{ij}$ in
$\kappa_{ij}L_{i}L_{j}HH$ is symmetric, i.e. it transforms under
the irreducible representation $\ydiagram{2}=\left\{ 2\right\} $
of $S_{2}$. This implies that there are $\mathcal{S}\left(\left\{ 2\right\} ,n\right)=n\left(n+1\right)/2$
operators of this type, so the second result (the permutation symmetry)
implies the first one (the operator counting). But does it contain
more information? For this particular example, the answer is no: there
is no extra information in the permutation symmetry because from the
number of operators as a function of $n$ it follows that $\kappa$
is necessarily a symmetric matrix.

More generally, we can frame the discussion as follows. The traditional
method provides the symmetry of the parameter tensors as a sum $\sum_{\lambda\vdash m}r_{\lambda}\lambda$
of irreducible representations $\lambda$ of some permutation group
$S_{m}$, each with multiplicity $r_{\lambda}$. From here one can
always calculate the total number of operators as a function of the
number of flavors $n$,
\begin{equation}
\#\textrm{operators}\left(n\right)=\sum_{\lambda\vdash m}r_{\lambda}\mathcal{S}\left(\lambda,n\right)\,,\label{eq:Information-comparison}
\end{equation}
while the Hilbert series method provides only the function $\#\textrm{operators}\left(n\right)$.\footnote{The number of flavors $n$ should be seen as a variable here, otherwise
$\#\textrm{operators}\left(n\right)$ becomes just a number and some
information given by the Hilbert series method is lost.} Knowing the left-hand-side of this expression, is it possible to
retrieve the integer coefficients $r_{\lambda}$? In simple cases
it is. The quantity $\mathcal{S}\left(\lambda,n\right)$ is a polynomial
function of $n$ of degree $m$ (because $\lambda$ is a representation
of $S_{m}$, hence it is a partition of $m$), with no constant term,
which we can express as 
\begin{equation}
\mathcal{S}\left(\lambda,n\right)\equiv\sum_{i=1}^{m}S_{\lambda i}n^{i}\,.
\end{equation}
We can obtain the $r_{\lambda}$ coefficients from $\#\textrm{operators}\left(n\right)$
in equation (\ref{eq:Information-comparison}) if and only if all
the polynomials $\mathcal{S}\left(\lambda,n\right)$ for different
partitions $\lambda$ of $m$ are linearly independent. In turn, this
is the same as requiring that all rows of the $S$ matrix above are
linearly independent. With either two or three repeated fields ($m=2$
or $3$), the expressions
\begin{align}
\mathcal{S}\left(\left\{ 2\right\} ,n\right) & =\frac{n\left(n+1\right)}{2}\,,\\
\mathcal{S}\left(\left\{ 1,1\right\} ,n\right) & =\frac{n\left(n-1\right)}{2}\,,\\
\mathcal{S}\left(\left\{ 3\right\} ,n\right) & =\frac{\left(n+2\right)\left(n+1\right)n}{6}\,,\\
\mathcal{S}\left(\left\{ 2,1\right\} ,n\right) & =\frac{\left(n+1\right)n\left(n-1\right)}{3}\,,\\
\mathcal{S}\left(\left\{ 1,1,1\right\} ,n\right) & =\frac{n\left(n-1\right)\left(n-2\right)}{6}\,,
\end{align}
imply that equation (\ref{eq:Information-comparison}) is reversible,
since the matrices
\begin{equation}
S=\left(\begin{array}{cc}
\frac{1}{2} & \frac{1}{2}\\
-\frac{1}{2} & \frac{1}{2}
\end{array}\right)_{m=2},\left(\begin{array}{ccc}
\frac{1}{3} & \frac{1}{2} & \frac{1}{6}\\
-\frac{1}{3} & 0 & \frac{1}{3}\\
\frac{1}{3} & -\frac{1}{2} & \frac{1}{6}
\end{array}\right)_{m=3}
\end{equation}
are invertible. Therefore, whenever a field appears repeated at most
3 times in an interaction, the Hilbert series method provides the
same information as the traditional method. But for $m>3$ the rows
of matrix $S$ are never linearly independent: this can be seen just
by counting its columns ($m$) and its rows (equal in number to the
distinct ways $p(m)$ of partitioning $m$). The function $p(m)$
(called the partition function) becomes bigger than $m$ starting
at $m=4$, increasing almost exponentially for very large $m$, so
for 4 or more fields, the traditional method does provide some extra
information which is not obtainable with the Hilbert series method
described in the literature. For example, there is one linear relation
among the $\mathcal{S}\left(\lambda,n\right)$ for $m=4$, namely
\begin{equation}
\mathcal{S}\left(\left\{ 3,1\right\} ,n\right)+\mathcal{S}\left(\left\{ 2,1,1\right\} ,n\right)=3\mathcal{S}\left(\left\{ 2,2\right\} ,n\right)\quad\left[=n^{2}\left(n^{2}-1\right)/4\right]\textrm{ }\,.
\end{equation}
Therefore, from the fact that there are $n^{2}\left(n^{2}-1\right)/4$
operators of type $d^{c}d^{c}d^{c}d^{c}\left(d^{c}\right)^{*}\left(e^{c}\right)^{*}$
in SMEFT it is impossible to infer what is the $S_{4}$ symmetry of
the 4 $d^{c}$'s; it might be $3\left\{ 2,2\right\} $ or $\left\{ 3,1\right\} +\left\{ 2,1,1\right\} $
(it turns out to be the latter case, so one Lagrangian term is enough
to write down all these operators). This is the smallest SMEFT operator
where the issue arises, and it has dimension 9, so it is fair to say
that the extra information given by the traditional method is either
minimal or non-existent at all, unless the dimension of the operators
is fairly large.

\subsection{Counting terms with derivatives}

Another important point worth mentioning is that the permutation symmetry
of operators with derivatives is unclear, due to the integration-by-part
redundancies. Consider the example in table \ref{tab:example-singlet}
with the operators of the kind $\partial^{4}S^{4}$, where $S$ is
a scalar field which does not change under gauge transformations.
There are operators with symmetries $S_{2}\times S_{2}$, $S_{2}\times S_{1}\times S_{1}$
and $S_{4}$ depending on where the derivatives are applied (these
are the operators with 0 $\mathcal{D}$'s in the table), and one must
cross out some of them due to redundancies which have other permutation
symmetries (they are associated with the operators with one or more
$\mathcal{D}$'s). It is hard to access what is the overall permutation
symmetry of what remains after this subtraction. In fact, just as
with the Hilbert series method, it is not even clear if any of the
three types of operators --- $SS\left(\partial^{2}S\right)\left(\partial^{2}S\right)$,
$S\left(\partial S\right)\left(\partial S\right)\left(\partial^{2}S\right)$
and $\left(\partial S\right)\left(\partial S\right)\left(\partial S\right)\left(\partial S\right)$
--- can be completely crossed out, hence one cannot tell how to distribute
the 4 derivatives over the 4 $S$ scalars. At least not without extra
considerations.

In light of this complication, one might simply compute the total
number of operators and ignore all other information since one cannot
make good use of it. However, there is perhaps a better alternative
where less information is thrown away. For simplicity, consider the
operators with only 2 derivatives and $m$ scalar singlets $S$: there
are $\left(\partial S\right)^{2}S^{m-2}$ operators, with a symmetry
$\left\{ 2\right\} _{\left(\partial S\right)}\times\left\{ m-2\right\} _{\left(S\right)}$,
and redundancies $\mathcal{D}\left[\left(\partial S\right)S^{m-1}\right]$
with a symmetry $\left\{ 1\right\} _{\left(\partial S\right)}\times\left\{ m-1\right\} _{\left(S\right)}$.
So, we need an $m$-index parameter tensor $\kappa$ to multiply/contract
with the $\left(\partial S\right)^{2}S^{m-2}$ operators, and it must
have a $\left\{ 2\right\} \times\left\{ m-2\right\} $ permutation
symmetry, with those components transforming as $\left\{ 1\right\} \times\left\{ m-1\right\} $
removed. The number of non-redundant operators must be
\begin{equation}
\mathcal{S}\left(\left\{ m-2\right\} ,n\right)\mathcal{S}\left(\left\{ 2\right\} ,n\right)-\mathcal{S}\left(\left\{ m-1\right\} ,n\right)\mathcal{S}\left(\left\{ 1\right\} ,n\right)=\begin{cases}
0 & \textrm{ if }n=1\\
\frac{n\left(m-3\right)\left(m+n-3\right)!}{2\left(m-1\right)!\left(n-2\right)!} & \textrm{ if }n>1
\end{cases}\,.\label{eq:87}
\end{equation}
However we can go beyond this counting exercise and, roughly speaking,
subtract the $\left\{ 1\right\} \times\left\{ m-1\right\} $ representation
of $S_{1}\times S_{m-1}$ from the $\left\{ 2\right\} \times\left\{ m-2\right\} $
representation of $S_{2}\times S_{m-2}$. Notice that these two representations
can be found inside the following $S_{m}$ representations exactly
once in all cases:
\begin{align}
\left\{ 2\right\} \times\left\{ m-2\right\}  & \subset\left\{ m\right\} ,\left\{ m-1,1\right\} ,\textrm{ and }\left\{ m-2,2\right\} ;\\
\left\{ 1\right\} \times\left\{ m-1\right\}  & \subset\left\{ m\right\} \textrm{ and }\left\{ m-1,1\right\} \,.
\end{align}
Therefore, only the components of $\kappa$ transforming as $\left\{ m-2,2\right\} $
under $S_{m}$ permutations survive.\footnote{Not all of them though: only those which transform as $\left\{ 2\right\} \times\left\{ m-2\right\} $
under $S_{2}\times S_{m-2}$ permutations. From the comments made
in section \ref{sec:3} one can infer that this corresponds exactly
to 1 in $d\left(\left\{ m-2,2\right\} \right)=m\left(m-3\right)/2$
components of $\kappa$ transforming as $\left\{ m-2,2\right\} $
under $S_{m}$ permutations.} And indeed it is straightforward to check that the result in equation
(\ref{eq:87}) is the same as $\mathcal{S}\left(\left\{ m-2,2\right\} ,n\right)$,
using formula (\ref{eq:S}).

In this way, even in the presence of derivatives, it is still possible
to obtain some permutation symmetry information of the operators with
the redundancies removed. For example, in the more complicated case
of operators of the kind $\partial^{4}S^{4}$ (see table \ref{tab:example-singlet})
it turns out that the non-redundant operators are associated with the
$S_{4}$ representation $\left\{ 4\right\} +\left\{ 2,2\right\} $,
and indeed the total number of non-redundant operators in this case
is
\begin{equation}
\mathcal{S}\left(\left\{ 4\right\} ,n\right)+\mathcal{S}\left(\left\{ 2,2\right\} ,n\right)=\frac{1}{8}n\left(n^{3}+2n^{2}+3n+2\right)\,.
\end{equation}

~

Finally, it is worthwhile to discuss the number of \textit{terms} [as defined
in section (\ref{sec:2})] associated with operators with derivatives.
In the absence of derivatives, the symmetry of the contraction of
the fields can be expressed as a sum $\sum_{\lambda}r_{\lambda}\lambda$
of irreducible representations $\lambda$ of the relevant permutation
group, with $r_{\lambda}$ representing the multiplicity of $\lambda$.
We have seen already that these operators can be expressed with $t=\max\left(r_{\lambda}\right)$
terms, and no less.

Integration-by-parts redundancies complicate this calculation for
operators with derivatives. Nevertheless, one can easily establish
bounds on the minimal number of terms $t$:
\begin{itemize}
\item $t$ cannot be smaller than the number of \textit{operators} when
considering only one generation of fields ($n=1$).
\item $t$ does not need to be larger than the number of irreducible representations
of the permutation group obtained after the procedure described a
few paragraphs earlier.
\item $t$ also cannot exceed the number of terms needed to write all interactions
if integration-by-parts redundancies are ignored.
\end{itemize}
For example, there is only one operator of the kind $\partial^{4}S^{4}$
if there is only one flavor of $S$, so $t\geq1$. On the other hand,
ignoring the redundancies in table \ref{tab:example-singlet}, one
could write all operators with 3 terms, so $t\leq3$. Finally, the
$\partial^{4}S^{4}$ interactions are associated with the $\left\{ 4\right\} +\left\{ 2,2\right\} $
permutation symmetry (two irreducible components), and therefore $t\leq2$.
In summary, these operators require either one or two Lagrangian terms
($1\leq t\leq2$).

\subsection{Application to specific models}

The approach described in this work can be used to characterize individual
interactions of a model. However, it would not be instructive to present
here an exhaustive analysis of this kind. Instead, I will show some
summary data of the interactions up to dimension 15 of three models:
(a) the SMEFT; (b) an $SU(5)$ model with the left-handed fermion
representations $3\times\overline{\boldsymbol{5}}+3\times\boldsymbol{10}$
and a scalar transforming as a $\boldsymbol{5}$; (c) an $SO(10)$
model with the left-handed fermion representations $3\times\boldsymbol{16}$
and a real scalar transforming as a $\boldsymbol{10}$. These two
latter models were picked to illustrate the effect of an enlarged
symmetry group on the number of interactions, so the only fermion
and scalar representations of $SU(5)$ and $SO(10)$ which were selected
are those which contain the Standard Model fields.

The total number of real operators, terms, and types of operators
up to dimension $d=2,\cdots,15$ in these three models is represented
graphically in figures \ref{fig:SMEFT} and \ref{fig:SMEFT-SU5-SO10}.
Exact numbers can found in appendix. In the case of SMEFT, the number
of operators up to dimension 15, as well as the number of types of
operators up to dimension 12 agree with the results obtained with
the Hilbert series method in \cite{Henning:2015alf}.

\begin{figure}[tbph]
\begin{centering}
\includegraphics[scale=0.5]{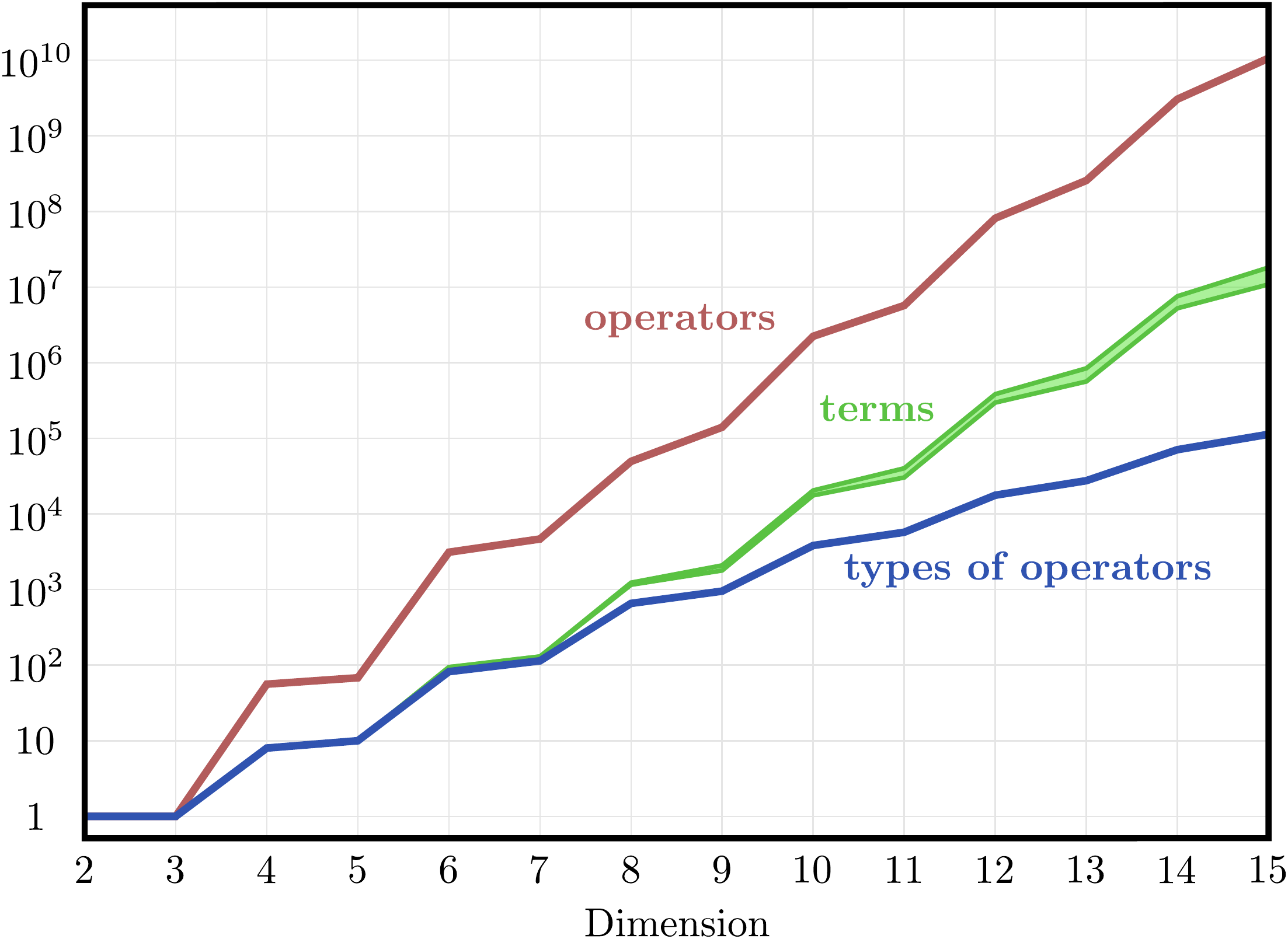}
\par\end{centering}
\caption{\label{fig:SMEFT}Cumulative number of real operators, terms, and
types of operators in SMEFT (as defined in section \ref{sec:2}) up
to a given dimension. Kinetics terms are ignored. Due to the presence
of derivatives in some operators, the number of required Lagrangian
terms cannot be determined exactly, but fairly stringent bounds (narrow
green band) can be set on this quantity.}
\end{figure}

As mentioned earlier, the minimum number of terms which are needed
to write down all operators can be calculated exactly for terms without
derivatives. If there are derivatives, with the considerations made
in this work it is only possible to derive bounds on this number.
One can see from the figures \ref{fig:SMEFT} and \ref{fig:SMEFT-SU5-SO10}
that these bounds are fairly strict. Furthermore, the lower limit
is close to (but never below) the number of operators in each of the
models if they had a single fermion family.

\begin{figure}[tbph]
\begin{centering}
\includegraphics[scale=0.5]{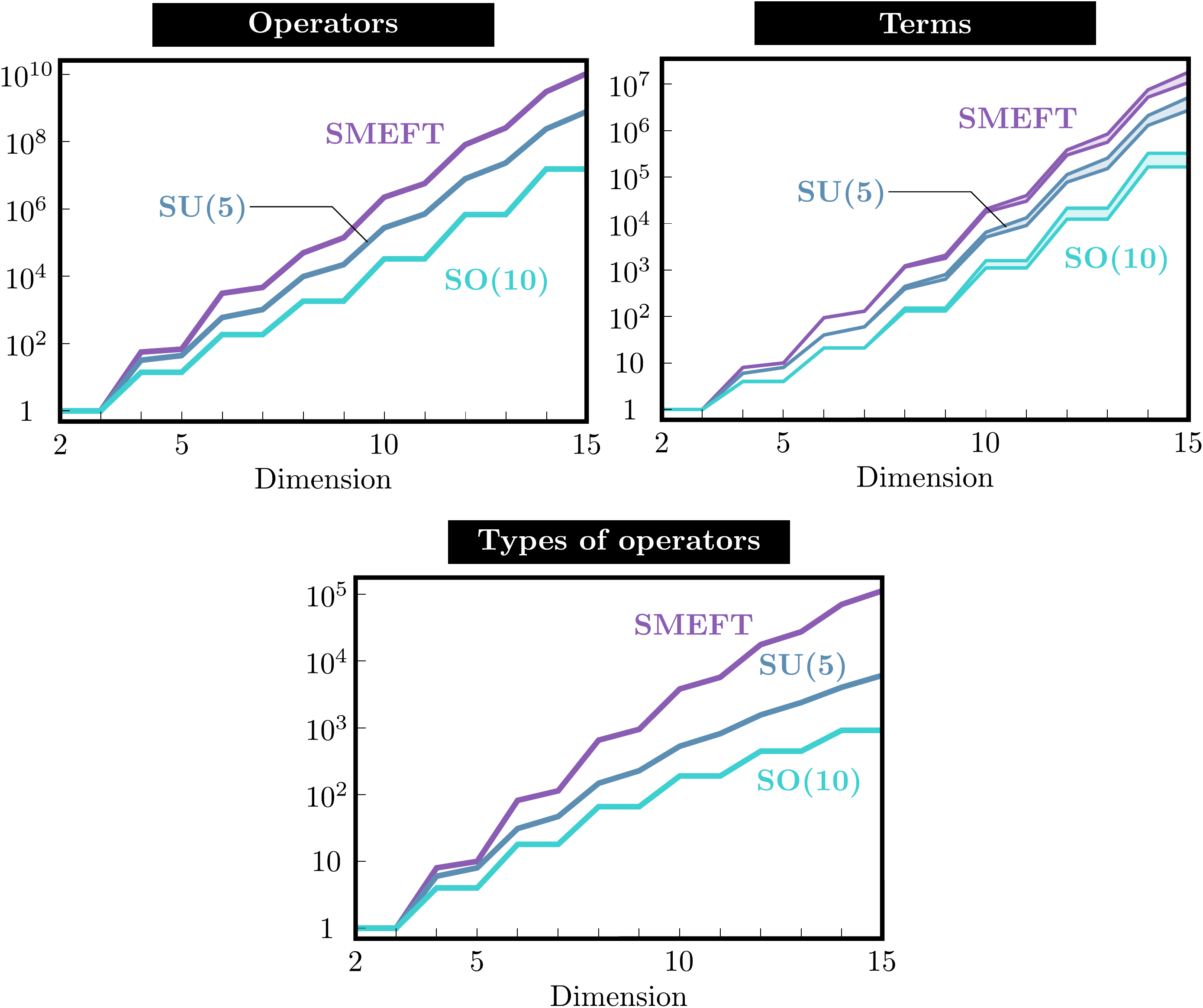}
\par\end{centering}
\caption{\label{fig:SMEFT-SU5-SO10}Cumulative number of real operators, terms,
and types of operators up to a given dimension in three models with
different gauge symmetries --- $SU(3)\times SU(2)\times U(1)$, $SU(5)$
and $SO(10)$. The fermion and scalar content of each model is
analogous (details can be found in the main text). As the symmetry
group is enlarged, the number of operators, terms, and types of operators
is reduced, as expected.}
\end{figure}

\section{\label{sec:Summary}Summary}

Operators in an effective field theory can be counted with the Hilbert
series method. This technique, which requires the computation of some integrals, 
is very different from the one historically used
in Particle Physics of simply multiplying together a model's fields in all possible
ways, and retaining the combinations which are invariant under all
relevant symmetries (such as the ones associated with the Lorentz and
gauge groups). However, such seemingly straightforward approach ---
which we may call the traditional method --- is complicated by the
fact that many operators have repeated fields, as well as derivatives.

This work describes a systematic and efficient way of addressing the
problem of repeated fields. Derivatives can be handled with simple
adaptations of the techniques known to work for the Hilbert series
method. With these two difficulties solved, it becomes possible to
count all operators of an effective field theory up to a high mass
dimension, without relying on the Hilbert series technique.

The traditional method is potentially useful for various reasons. An
obvious one is that it makes it possible to cross-check results obtained
in other ways. For example, the operators of the Standard Model effective
field theory (SMEFT) were previously counted up to dimension 15 with
the Hilbert series technique, but in the literature there was only
confirmation of these numbers up to dimension 8. With the procedure
described in this work, it is possible to verify the number of operators
up to dimension 15, for an arbitrary number of fermion families.

Another interesting feature of the traditional method is that it provides
more information beyond a simple operator counting. Among other things,
the extra information can be used to count systematically the minimum
number of Lagrangian terms required to encode a model's interactions.
For instance, it indicates that all $QQQL$ operators in SMEFT can
be written down as a single term in the Lagrangian (in the past, this
was not always clear), while the significantly more complicated interactions
with twice as many fields, of the type $Q^{6}L^{2}$, need 11 terms.

The method described in this work was implemented in the program \texttt{Sym2Int},
which can readily be used to calculate the above operator properties
in any field theory.

\section*{Acknowledgments}

I would like to thank Xiaochuan Lu and Tom Melia for helping me understand
one of their calculations in reference \cite{Henning:2015alf}.
I acknowledge the financial support from the Grant Agency of the
Czech Republic (GA\v{C}R) through contract number 17-04902S, from the
Charles University Research Center UNCE/SCI/013, and from
the Generalitat Valenciana through the grant SEJI/2018/033.

\clearpage{}

\section*{\label{sec:Appendix}Appendix\addcontentsline{toc}{section}{\protect\numberline{}Appendix}}

The exact number of operators, terms and types of operators for the
three models described in section \ref{sec:5} are given here in tables
\ref{tab:SMEFT}, \ref{tab:SU5} and \ref{tab:SO10}. Kinetic terms
were not taken into account. Furthermore, note that any complex operator
(plus its Hermitian conjugate) can be written as two real ones, and
vice-versa; the numbers in the tables below refer to real operators,
terms, and term types.

\begin{table}[H]
\begin{centering}
\begin{tabular}{cccc}
\toprule 
Dimension & \# operators & \# terms & \# types of operators\tabularnewline
\midrule
2 & 1 & 1 & 1\tabularnewline
3 & 0 & 0 & 0\tabularnewline
4 & 55 & 7 & 7\tabularnewline
5 & 12 & 2 & 2\tabularnewline
6 & 3045 & 84 & 72\tabularnewline
7 & 1542 & 36 & 32\tabularnewline
8 & 44807 & 1025 to 1102 & 541\tabularnewline
9 & 90456 & 628 to 852 & 296\tabularnewline
10 & 2092441 & 15769 to 18345 & 2868\tabularnewline
11 & 3472266 & 12726 to 19666 & 1898\tabularnewline
12 & 75577476 & 266031 to 343511 & 11942\tabularnewline
13 & 175373592 & 266802 to 457898 & 9824\tabularnewline
14 & 2795173575 & 4669533 to 6717444 & 43158\tabularnewline
15 & 7557369962 & 5599846 to 10567408 & 42206\tabularnewline
\bottomrule
\end{tabular}
\par\end{centering}
\caption{\label{tab:SMEFT}Number of real operators, terms, and types of operators
in SMEFT (as defined in section \ref{sec:2}). Kinetics terms ---
of the kind $\partial\partial\phi\phi$, $\partial\psi^{*}\psi$ and
$FF$ --- are not taken into account. The number of operators up
to dimension 15 shown here, as well as the number of types of operators
up to dimension 12 agree with \cite{Henning:2015alf}.}
\end{table}
\begin{table}[tbh]
\begin{centering}
\begin{tabular}{cccc}
\toprule 
Dimension & \# operators & \# terms & \# types of operators\tabularnewline
\midrule
2 & 1 & 1 & 1\tabularnewline
3 & 0 & 0 & 0\tabularnewline
4 & 31 & 5 & 5\tabularnewline
5 & 12 & 2 & 2\tabularnewline
6 & 549 & 32 & 23\tabularnewline
7 & 432 & 20 & 16\tabularnewline
8 & 8761 & 332 to 387 & 101\tabularnewline
9 & 12392 & 242 to 352 & 80\tabularnewline
10 & 252626 & 4459 to 5720 & 302\tabularnewline
11 & 431670 & 3990 to 6770 & 288\tabularnewline
12 & 7159650 & 68577 to 99666 & 743\tabularnewline
13 & 15425382 & 74252 to 142748 & 834\tabularnewline
14 & 215670831 & 1139503 to 1850054 & 1651\tabularnewline
15 & 544121758 & 1449420 to 3067252 & 2056\tabularnewline
\bottomrule
\end{tabular}
\par\end{centering}
\caption{\label{tab:SU5}Number of real operators, terms, and types of operators
in an $SU(5)$ model with the left-handed fermion representations
$3\times\overline{\boldsymbol{5}}+3\times\boldsymbol{10}$ and a scalar
transforming as a $\boldsymbol{5}$.}
\end{table}

The reader will notice from table \ref{tab:SO10} that the $SO(10)$
model with left-handed fermions in the spinor representation $\boldsymbol{16}$
and a scalar in the vector representation $\boldsymbol{10}$ does
not seem to have operators with an odd dimension. It can be shown
analytically that this is indeed the case. Let us call $n_{16}$,
$n_{16^{*}}$, $n_{10}$ and $n_{D}$ to the number of $\boldsymbol{16}$'s,
$\boldsymbol{16}^{*}$'s, $\boldsymbol{10}$'s and derivatives $D$
in a given interaction (for all practical purposes, here $\mathcal{F}=D^{2}$).
Then Lorentz invariance requires that $n_{16}+n_{D}$ and $n_{16^{*}}+n_{D}$
are both even numbers, while from the conjugacy classes of the $SO(10)$
representations we also conclude that $n_{16}+n_{16^{*}}$ must be
even, and $2n_{10}+n_{16}-n_{16^{*}}$ must be a multiple of 4. It
follows that the operator dimension 
\begin{equation}
d=\frac{3}{2}n_{16}+\frac{3}{2}n_{16^{*}}+n_{10}+n_{D}=\left(n_{16}+n_{16^{*}}\right)+\left(n_{16^{*}}+n_{D}\right)+\frac{1}{2}\left(2n_{10}+n_{16}-n_{16^{*}}\right)
\end{equation}
must be even.
\begin{table}[tbph]
\begin{centering}
\begin{tabular}{cccc}
\toprule 
Dimension & \# operators & \# terms & \# types of operators\tabularnewline
\midrule
2 & 1 & 1 & 1\tabularnewline
3 & 0 & 0 & 0\tabularnewline
4 & 13 & 3 & 3\tabularnewline
5 & 0 & 0 & 0\tabularnewline
6 & 170 & 17 & 14\tabularnewline
7 & 0 & 0 & 0\tabularnewline
8 & 1639 & 110 to 131 & 48\tabularnewline
9 & 0 & 0 & 0\tabularnewline
10 & 31059 & 977 to 1440 & 124\tabularnewline
11 & 0 & 0 & 0\tabularnewline
12 & 648654 & 11319 to 19765 & 257\tabularnewline
13 & 0 & 0 & 0\tabularnewline
14 & 14694065 & 152402 to 302812 & 472\tabularnewline
15 & 0 & 0 & 0\tabularnewline
\bottomrule
\end{tabular}
\par\end{centering}
\caption{\label{tab:SO10}Number of real operators, terms, and types of operators
in an $SO(10)$ model with the left-handed fermion representations
$3\times\boldsymbol{16}$ and a real scalar transforming as a $\boldsymbol{10}$.
It can be shown that such a model does not have operators with an
odd dimension.}
\end{table}

\clearpage{}


\begin{thebibliography}{10}
	\providecommand{\url}[1]{\texttt{#1}}
	\providecommand{\urlprefix}{URL }
	\providecommand{\eprint}[2][]{\url{#2}}
	%% \vspace{-0.1em}
	\bibitem{Weinberg:1979sa}
	S.~Weinberg, \emph{{Baryon and lepton nonconserving processes}},
	\MYhref[journalLinks]{http://dx.doi.org/10.1103/PhysRevLett.43.1566}{Phys.
		Rev. Lett.
	}\MYhref[journalLinks]{http://dx.doi.org/10.1103/PhysRevLett.43.1566}{\textbf{43}
		(1979) 1566--1570}.
	
	\bibitem{Wilczek:1979hc}
	F.~Wilczek and A.~Zee, \emph{{Operator analysis of nucleon decay}},
	\MYhref[journalLinks]{http://dx.doi.org/10.1103/PhysRevLett.43.1571}{Phys.
		Rev. Lett.
	}\MYhref[journalLinks]{http://dx.doi.org/10.1103/PhysRevLett.43.1571}{\textbf{43}
		(1979) 1571--1573}.
	
	\bibitem{Abbott:1980zj}
	L.~F. Abbott and M.~B. Wise, \emph{{The effective Hamiltonian for nucleon
			decay}},
	\MYhref[journalLinks]{http://dx.doi.org/10.1103/PhysRevD.22.2208}{Phys. Rev.
	}\MYhref[journalLinks]{http://dx.doi.org/10.1103/PhysRevD.22.2208}{\textbf{D22}
		(1980) 2208}.
	
	\bibitem{Burges:1983zg}
	C.~J.~C. Burges and H.~J. Schnitzer, \emph{{Virtual effects of excited quarks
			as probes of a possible new hadronic mass scale}},
	\MYhref[journalLinks]{http://dx.doi.org/10.1016/0550-3213(83)90555-2}{Nucl.
		Phys.
	}\MYhref[journalLinks]{http://dx.doi.org/10.1016/0550-3213(83)90555-2}{\textbf{B228}
		(1983) 464--500}.
	
	\bibitem{Leung:1984ni}
	C.~N. Leung, S.~T. Love and S.~Rao, \emph{{Low-energy manifestations of a new
			interaction scale: operator analysis}},
	\MYhref[journalLinks]{http://dx.doi.org/10.1007/BF01588041}{Z. Phys.
	}\MYhref[journalLinks]{http://dx.doi.org/10.1007/BF01588041}{\textbf{C31}
		(1986) 433}.
	
	\bibitem{Buchmuller:1985jz}
	W.~Buchm\"uller and D.~Wyler, \emph{{Effective Lagrangian analysis of new
			interactions and flavor conservation}},
	\MYhref[journalLinks]{http://dx.doi.org/10.1016/0550-3213(86)90262-2}{Nucl.
		Phys.
	}\MYhref[journalLinks]{http://dx.doi.org/10.1016/0550-3213(86)90262-2}{\textbf{B268}
		(1986) 621--653}.
	
	\bibitem{Grzadkowski:2010es}
	B.~Grzadkowski, M.~Iskrzy\'nski, M.~Misiak and J.~Rosiek, \emph{{Dimension-six
			terms in the Standard Model Lagrangian}},
	\MYhref[journalLinks]{http://dx.doi.org/10.1007/JHEP10(2010)085}{JHEP
	}\MYhref[journalLinks]{http://dx.doi.org/10.1007/JHEP10(2010)085}{\textbf{10}
		(2010) 085}, \MYhref[eprintLinks]{http://arxiv.org/abs/1008.4884}{{\ttfamily
			arXiv:1008.4884 [hep-ph]}}.
	
	\bibitem{Alonso:2013hga}
	R.~Alonso, E.~E. Jenkins, A.~V. Manohar and M.~Trott, \emph{{Renormalization
			group evolution of the Standard Model dimension six operators III: gauge
			coupling dependence and phenomenology}},
	\MYhref[journalLinks]{http://dx.doi.org/10.1007/JHEP04(2014)159}{JHEP
	}\MYhref[journalLinks]{http://dx.doi.org/10.1007/JHEP04(2014)159}{\textbf{04}
		(2014) 159}, \MYhref[eprintLinks]{http://arxiv.org/abs/1312.2014}{{\ttfamily
			arXiv:1312.2014 [hep-ph]}}.
	
	\bibitem{Lehman:2014jma}
	L.~Lehman, \emph{{Extending the Standard Model effective field theory with the
			complete set of dimension-7 operators}},
	\MYhref[journalLinks]{http://dx.doi.org/10.1103/PhysRevD.90.125023}{Phys.
		Rev.
	}\MYhref[journalLinks]{http://dx.doi.org/10.1103/PhysRevD.90.125023}{\textbf{D90}
		(2014) 12 125023},
	\MYhref[eprintLinks]{http://arxiv.org/abs/1410.4193}{{\ttfamily
			arXiv:1410.4193 [hep-ph]}}.
	
	\bibitem{Liao:2016hru}
	Y.~Liao and X.-D. Ma, \emph{{Renormalization group evolution of dimension-seven
			baryon- and lepton-number-violating operators}},
	\MYhref[journalLinks]{http://dx.doi.org/10.1007/JHEP11(2016)043}{JHEP
	}\MYhref[journalLinks]{http://dx.doi.org/10.1007/JHEP11(2016)043}{\textbf{11}
		(2016) 043}, \MYhref[eprintLinks]{http://arxiv.org/abs/1607.07309}{{\ttfamily
			arXiv:1607.07309 [hep-ph]}}.
	
	\bibitem{Liao:2019tep}
	Y.~Liao and X.-D. Ma, \emph{{Renormalization group evolution of dimension-seven
			operators in Standard Model effective field theory and relevant
			phenomenology}},
	\MYhref[journalLinks]{http://dx.doi.org/10.1007/JHEP03(2019)179}{JHEP
	}\MYhref[journalLinks]{http://dx.doi.org/10.1007/JHEP03(2019)179}{\textbf{03}
		(2019) 179}, \MYhref[eprintLinks]{http://arxiv.org/abs/1901.10302}{{\ttfamily
			arXiv:1901.10302 [hep-ph]}}.
	
	\bibitem{Babu:2001ex}
	K.~S. Babu and C.~N. Leung, \emph{{Classification of effective neutrino mass
			operators}},
	\MYhref[journalLinks]{http://dx.doi.org/10.1016/S0550-3213(01)00504-1}{Nucl.
		Phys.
	}\MYhref[journalLinks]{http://dx.doi.org/10.1016/S0550-3213(01)00504-1}{\textbf{B619}
		(2001) 667--689},
	\MYhref[eprintLinks]{http://arxiv.org/abs/hep-ph/0106054}{{\ttfamily
			arXiv:hep-ph/0106054 [hep-ph]}}.
	
	\bibitem{deGouvea:2007qla}
	A.~de~Gouv\^ea and J.~Jenkins, \emph{{A survey of lepton number violation via
			effective operators}},
	\MYhref[journalLinks]{http://dx.doi.org/10.1103/PhysRevD.77.013008}{Phys.
		Rev.
	}\MYhref[journalLinks]{http://dx.doi.org/10.1103/PhysRevD.77.013008}{\textbf{D77}
		(2008) 013008},
	\MYhref[eprintLinks]{http://arxiv.org/abs/0708.1344}{{\ttfamily
			arXiv:0708.1344 [hep-ph]}}.
	
	\bibitem{Fonseca:2018aav}
	R.~M. Fonseca and M.~Hirsch, \emph{{$\Delta L \ge 4$ lepton number violating
			processes}},
	\MYhref[journalLinks]{http://dx.doi.org/10.1103/PhysRevD.98.015035}{Phys.
		Rev.
	}\MYhref[journalLinks]{http://dx.doi.org/10.1103/PhysRevD.98.015035}{\textbf{D98}
		(2018) 1 015035},
	\MYhref[eprintLinks]{http://arxiv.org/abs/1804.10545}{{\ttfamily
			arXiv:1804.10545 [hep-ph]}}.
	
	\bibitem{Lehman:2015via}
	L.~Lehman and A.~Martin, \emph{{Hilbert series for constructing Lagrangians:
			expanding the phenomenologist's toolbox}},
	\MYhref[journalLinks]{http://dx.doi.org/10.1103/PhysRevD.91.105014}{Phys.
		Rev.
	}\MYhref[journalLinks]{http://dx.doi.org/10.1103/PhysRevD.91.105014}{\textbf{D91}
		(2015) 105014},
	\MYhref[eprintLinks]{http://arxiv.org/abs/1503.07537}{{\ttfamily
			arXiv:1503.07537 [hep-ph]}}.
	
	\bibitem{Lehman:2015coa}
	L.~Lehman and A.~Martin, \emph{{Low-derivative operators of the Standard Model
			effective field theory via Hilbert series methods}},
	\MYhref[journalLinks]{http://dx.doi.org/10.1007/JHEP02(2016)081}{JHEP
	}\MYhref[journalLinks]{http://dx.doi.org/10.1007/JHEP02(2016)081}{\textbf{02}
		(2016) 081}, \MYhref[eprintLinks]{http://arxiv.org/abs/1510.00372}{{\ttfamily
			arXiv:1510.00372 [hep-ph]}}.
	
	\bibitem{Henning:2015daa}
	B.~Henning, X.~Lu, T.~Melia and H.~Murayama, \emph{{Hilbert series and operator
			bases with derivatives in effective field theories}},
	\MYhref[journalLinks]{http://dx.doi.org/10.1007/s00220-015-2518-2}{Commun.
		Math. Phys.
	}\MYhref[journalLinks]{http://dx.doi.org/10.1007/s00220-015-2518-2}{\textbf{347}
		(2016) 2 363--388},
	\MYhref[eprintLinks]{http://arxiv.org/abs/1507.07240}{{\ttfamily
			arXiv:1507.07240 [hep-th]}}.
	
	\bibitem{Henning:2015alf}
	B.~Henning, X.~Lu, T.~Melia and H.~Murayama, \emph{{2, 84, 30, 993, 560, 15456,
			11962, 261485, ...: higher dimension operators in the SM EFT}},
	\MYhref[journalLinks]{http://dx.doi.org/10.1007/JHEP08(2017)016}{JHEP
	}\MYhref[journalLinks]{http://dx.doi.org/10.1007/JHEP08(2017)016}{\textbf{08}
		(2017) 016}, \MYhref[eprintLinks]{http://arxiv.org/abs/1512.03433}{{\ttfamily
			arXiv:1512.03433 [hep-ph]}} [Erratum: \MYhref[journalLinks]{http://doi.org/10.1007/JHEP09(2019)019}{JHEP 09, 019 (2019)}].
		
	\bibitem{Benvenuti:2006qr}
	S.~Benvenuti, B.~Feng, A.~Hanany and Y.-H. He, \emph{{Counting BPS operators in
			gauge theories: quivers, syzygies and plethystics}},
	\MYhref[journalLinks]{http://dx.doi.org/10.1088/1126-6708/2007/11/050}{JHEP
	}\MYhref[journalLinks]{http://dx.doi.org/10.1088/1126-6708/2007/11/050}{\textbf{11}
		(2007) 050},
	\MYhref[eprintLinks]{http://arxiv.org/abs/hep-th/0608050}{{\ttfamily
			arXiv:hep-th/0608050 [hep-th]}}.
	
	\bibitem{Feng:2007ur}
	B.~Feng, A.~Hanany and Y.-H. He, \emph{{Counting gauge invariants: The
			Plethystic program}},
	\MYhref[journalLinks]{http://dx.doi.org/10.1088/1126-6708/2007/03/090}{JHEP
	}\MYhref[journalLinks]{http://dx.doi.org/10.1088/1126-6708/2007/03/090}{\textbf{03}
		(2007) 090},
	\MYhref[eprintLinks]{http://arxiv.org/abs/hep-th/0701063}{{\ttfamily
			arXiv:hep-th/0701063 [hep-th]}}.
	
	\bibitem{Gray:2008yu}
	J.~Gray et~al., \emph{{SQCD: a geometric apercu}},
	\MYhref[journalLinks]{http://dx.doi.org/10.1088/1126-6708/2008/05/099}{JHEP
	}\MYhref[journalLinks]{http://dx.doi.org/10.1088/1126-6708/2008/05/099}{\textbf{05}
		(2008) 099}, \MYhref[eprintLinks]{http://arxiv.org/abs/0803.4257}{{\ttfamily
			arXiv:0803.4257 [hep-th]}}.
	
	\bibitem{Jenkins:2009dy}
	E.~E. Jenkins and A.~V. Manohar, \emph{{Algebraic structure of lepton and quark
			flavor invariants and CP violation}},
	\MYhref[journalLinks]{http://dx.doi.org/10.1088/1126-6708/2009/10/094}{JHEP
	}\MYhref[journalLinks]{http://dx.doi.org/10.1088/1126-6708/2009/10/094}{\textbf{10}
		(2009) 094}, \MYhref[eprintLinks]{http://arxiv.org/abs/0907.4763}{{\ttfamily
			arXiv:0907.4763 [hep-ph]}}.
	
	\bibitem{Hanany:2010vu}
	A.~Hanany, E.~E. Jenkins, A.~V. Manohar and G.~Torri, \emph{{Hilbert series for
			flavor invariants of the Standard Model}},
	\MYhref[journalLinks]{http://dx.doi.org/10.1007/JHEP03(2011)096}{JHEP
	}\MYhref[journalLinks]{http://dx.doi.org/10.1007/JHEP03(2011)096}{\textbf{03}
		(2011) 096}, \MYhref[eprintLinks]{http://arxiv.org/abs/1010.3161}{{\ttfamily
			arXiv:1010.3161 [hep-ph]}}.
	
	\bibitem{Bednyakov:2018cmx}
	A.~V. Bednyakov,
	\emph{{On three-loop RGE for the Higgs sector of 2HDM}}, \MYhref[journalLinks]{http://dx.doi.org/10.1007/JHEP11(2018)154}{JHEP \textbf{11} (2018) 154},
	\MYhref[eprintLinks]{http://arxiv.org/abs/1809.04527}{{\ttfamily
			arXiv:1809.04527 [hep-ph]}}.
	
	\bibitem{Trautner:2018ipq}
	A.~Trautner, \emph{{Systematic construction of basis invariants in the 2HDM}},
	\MYhref[journalLinks]{http://dx.doi.org/10.1007/JHEP05(2019)208}{JHEP
	}\MYhref[journalLinks]{http://dx.doi.org/10.1007/JHEP05(2019)208}{\textbf{05}
		(2019) 208}, \MYhref[eprintLinks]{http://arxiv.org/abs/1812.02614}{{\ttfamily
			arXiv:1812.02614 [hep-ph]}}.
	
	\bibitem{Anisha:2019nzx}
	Anisha, S.~Das~Bakshi, J.~Chakrabortty and S.~Prakash, \emph{{Hilbert series
			and plethystics: paving the path towards 2HDM- and MLRSM-EFT}},
	\MYhref[journalLinks]{http://dx.doi.org/10.1007/JHEP09(2019)035}{JHEP
	}\MYhref[journalLinks]{http://dx.doi.org/10.1007/JHEP09(2019)035}{\textbf{09}
		(2019) 035}, \MYhref[eprintLinks]{http://arxiv.org/abs/1905.11047}{{\ttfamily
			arXiv:1905.11047 [hep-ph]}}.
	
	\bibitem{Fonseca:2011sy}
	R.~M. Fonseca, \emph{{Calculating the renormalisation group equations of a SUSY
			model with Susyno}},
	\MYhref[journalLinks]{http://dx.doi.org/10.1016/j.cpc.2012.05.017}{Comput.
		Phys. Commun.
	}\MYhref[journalLinks]{http://dx.doi.org/10.1016/j.cpc.2012.05.017}{\textbf{183}
		(2012) 2298--2306},
	\MYhref[eprintLinks]{http://arxiv.org/abs/1106.5016}{{\ttfamily
			arXiv:1106.5016 [hep-ph]}}.
	
	\bibitem{Fonseca:2017lem}
	R.~M. Fonseca, \emph{{The Sym2Int program: going from symmetries to
			interactions}},
	\MYhref[journalLinks]{http://dx.doi.org/10.1088/1742-6596/873/1/012045}{J.
		Phys. Conf. Ser.
	}\MYhref[journalLinks]{http://dx.doi.org/10.1088/1742-6596/873/1/012045}{\textbf{873}
		(2017) 1 012045},
	\MYhref[eprintLinks]{http://arxiv.org/abs/1703.05221}{{\ttfamily
			arXiv:1703.05221 [hep-ph]}}.
	
	\bibitem{Slansky:1981yr}
	R.~Slansky, \emph{{Group theory for unified model building}},
	\MYhref[journalLinks]{http://dx.doi.org/10.1016/0370-1573(81)90092-2}{Phys.
		Rept.
	}\MYhref[journalLinks]{http://dx.doi.org/10.1016/0370-1573(81)90092-2}{\textbf{79}
		(1981) 1--128}.
	
	\bibitem{Cahn:1985wk}
	R.~N. Cahn, \emph{{Semi-simple Lie algebras and their representations}}, Dover Publications (1985).
	
	\bibitem{Cheng:1985bj}
	T.~P. Cheng and L.~F. Li, \emph{{Gauge theory of elementary particle physics}},
	 Oxford University Press (1984).
	
	\bibitem{Tung-book}
	W.~Tung, \emph{Group theory in Physics}, World Scientific (1985).

	\bibitem{Bernstein:2003}
	D.~{Bernstein}, \emph{The computational complexity of rules for the character
		table of $S_n$},
	\MYhref[journalLinks]{https://doi.org/10.1016/j.jsc.2003.11.001}{Journal
		of Symbolic Computation
	}\MYhref[journalLinks]{https://doi.org/10.1016/j.jsc.2003.11.001}{\textbf{37}
		(2004) 6 727 -- 748},
	\MYhref[eprintLinks]{http://arxiv.org/abs/math/0309225}{{\ttfamily
			arXiv:math/0309225 [math.CO]}}.
		
	\bibitem{Hook-length-formula-1954}
	J.~S. Frame, G.~d.~B. Robinson and R.~M. Thrall, \emph{The hook graphs of the
		symmetric group},
	\MYhref[journalLinks]{http://dx.doi.org/10.4153/CJM-1954-030-1}{Canadian
		Journal of Mathematics
	}\MYhref[journalLinks]{http://dx.doi.org/10.4153/CJM-1954-030-1}{\textbf{6}
		(1954) 316--324}.
	
	\bibitem{Stanley:1999}
	R.~P. Stanley, \emph{Enumerative combinatorics: volume 2}, Cambridge University
	Press (1999).
	
	\bibitem{Liao:2016qyd}
	Y.~Liao and X.-D. Ma, \emph{{Operators up to dimension seven in Standard Model
			effective field theory extended with sterile neutrinos}},
	\MYhref[journalLinks]{http://dx.doi.org/10.1103/PhysRevD.96.015012}{Phys.
		Rev.
	}\MYhref[journalLinks]{http://dx.doi.org/10.1103/PhysRevD.96.015012}{\textbf{D96}
		(2017) 1 015012},
	\MYhref[eprintLinks]{http://arxiv.org/abs/1612.04527}{{\ttfamily
			arXiv:1612.04527 [hep-ph]}}.
	
	\bibitem{Gripaios:2018zrz}
	B.~Gripaios and D.~Sutherland, \emph{{DEFT: a program for operators in EFT}},
	\MYhref[journalLinks]{http://dx.doi.org/10.1007/JHEP01(2019)128}{JHEP
	}\MYhref[journalLinks]{http://dx.doi.org/10.1007/JHEP01(2019)128}{\textbf{01}
		(2019) 128}, \MYhref[eprintLinks]{http://arxiv.org/abs/1807.07546}{{\ttfamily
			arXiv:1807.07546 [hep-ph]}}.
	
	\bibitem{Politzer:1980me}
	H.~D. Politzer, \emph{{Power corrections at short distances}},
	\MYhref[journalLinks]{http://dx.doi.org/10.1016/0550-3213(80)90172-8}{Nucl.
		Phys.
	}\MYhref[journalLinks]{http://dx.doi.org/10.1016/0550-3213(80)90172-8}{\textbf{B172}
		(1980) 349--382}.
	
	\bibitem{Arzt:1993gz}
	C.~Arzt, \emph{{Reduced effective Lagrangians}},
	\MYhref[journalLinks]{http://dx.doi.org/10.1016/0370-2693(94)01419-D}{Phys.
		Lett.
	}\MYhref[journalLinks]{http://dx.doi.org/10.1016/0370-2693(94)01419-D}{\textbf{B342}
		(1995) 189--195},
	\MYhref[eprintLinks]{http://arxiv.org/abs/hep-ph/9304230}{{\ttfamily
			arXiv:hep-ph/9304230 [hep-ph]}}.
	
	\bibitem{Henning:2017fpj}
	B.~Henning, X.~Lu, T.~Melia and H.~Murayama, \emph{{Operator bases,
			$S$-matrices, and their partition functions}},
	\MYhref[journalLinks]{http://dx.doi.org/10.1007/JHEP10(2017)199}{JHEP
	}\MYhref[journalLinks]{http://dx.doi.org/10.1007/JHEP10(2017)199}{\textbf{10}
		(2017) 199}, \MYhref[eprintLinks]{http://arxiv.org/abs/1706.08520}{{\ttfamily
			arXiv:1706.08520 [hep-th]}}.
	
	\bibitem{LieProgram}
	M.~A.~A. van Leeuwen, A.~M. Cohen and B.~Lisser, \emph{LiE, A package for Lie
		group computations} (1992), ISBN 90-74116-02-7,
	\url{http://wwwmathlabo.univ-poitiers.fr/~maavl/LiE/}.
	
	\bibitem{Fonseca:GroupMath}
	R. M. Fonseca, \emph{GroupMath: A Mathematica package for group theory calculations} (In preparation),
	\url{http://renatofonseca.net/groupmath.php}.
	
	\bibitem{Criado:2019ugp}
	J.~C. Criado, \emph{{BasisGen: automatic generation of operator bases}},
	\MYhref[journalLinks]{http://dx.doi.org/10.1140/epjc/s10052-019-6769-5}{Eur.
		Phys. J.
	}\MYhref[journalLinks]{http://dx.doi.org/10.1140/epjc/s10052-019-6769-5}{\textbf{C79}
		(2019) 3 256},
	\MYhref[eprintLinks]{http://arxiv.org/abs/1901.03501}{{\ttfamily
			arXiv:1901.03501 [hep-ph]}}.
	
\end{thebibliography}
\end{document}